# A recoiling supermassive black hole in a powerful quasar

**Marco Chiaberge*[1,2], Takahiro Morishita[3], Matteo Boschini[4,5], Stefano Bianchi[6], Alessandro Capetti[7], Gianluca Castignani[8], Davide Gerosa[4,5], Masahiro Konishi[9], Shuhei Koyama[9,10], Kosuke Kushibiki[13], Erini Lambrides[11], Eileen T. Meyer[12], Kentaro Motohara[9,13], Massimo Stiavelli[2,14,15], Hidenori Takahashi[16], Grant R. Tremblay[17], and Colin Norman[2,14]**

[1] Space Telescope Science Institute for the European Space Agency (ESA), ESA Office, 3700 San Martin Drive, Baltimore, MD, USA

[2] The William H. Miller III Department of Physics & Astronomy, Johns Hopkins University, Baltimore, MD, USA

[3] IPAC, California Institute of Technology, MC 314-6, 1200 E. California Boulevard, Pasadena, CA 91125, USA

[4] Dipartimento di Fisica "G. Occhialini", Università degli Studi di Milano-Bicocca, Piazza della Scienza 3, 20126 Milano, Italy

[5] INFN, Sezione di Milano-Bicocca, Piazza della Scienza 3, 20126 Milano, Italy

[6] Dipartimento di Matematica e Fisica, Università degli Studi Roma Tre, via della Vasca Navale 84, I-00146 Roma, Italy

[7] INAF - Osservatorio Astrofisico di Torino, Via Osservatorio 20, I-10025 Pino Torinese, Italy

[8] INAF-Osservatorio di Astrofisica e Scienza dello Spazio di Bologna, Via Piero Gobetti 93/3, 40129, Bologna, Italy

[9] Institute of Astronomy, Graduate School of Science, the University of Tokyo, Osawa 2-21-1, Mitaka, Tokyo 181-0015, Japan

[10] Astronomy Data Center, National Astronomical Observatory of Japan, Osawa 2-21-1, Mitaka, Tokyo 181-8588, Japan

[11] NASA-Goddard Space Flight Center, Code 662, Greenbelt, MD 20771, USA

[12] Department of Physics, University of Maryland, Baltimore County, Baltimore, MD, USA

[13] Advanced Technology Center, National Astronomical Observatory of Japan, Osawa 2-21-1, Mitaka, Tokyo 181-8588, Japan

[14] Space Telescope Science Institute, 3700 San Martin Drive, Baltimore, MD, USA

[15] Dept. of Astronomy, University of Maryland, College Park, MD 20742, USA

[16] Kiso Observatory, Institute of Astronomy, School of Science, the University of Tokyo, 10762-30 Mitake, Kiso, Nagano 397-0101, Japan

[17] Center for Astrophysics | Harvard & Smithsonian, 60 Garden Street, Cambridge, MA 02138, USA

*Corresponding author. Email: marcoc@stsci.edu

**ABSTRACT**

Supermassive black holes (SMBH) are thought to grow through accretion of matter and mergers. Models of SMBH mergers have long suffered the *final parsec problem*, where SMBH binaries may stall before energy loss from gravitational waves (GW) becomes significant, leaving the pair unmerged. Direct evidence of coalesced SMBH remains elusive. Theory predicts that GW recoiling black holes can occur following a black hole merger. Here we present new and conclusive spectroscopic evidence that both the accretion disk and the broad line region in the spatially offset quasar 3C 186 are blue-shifted by the same velocity relative to the host galaxy, with a line of sight velocity of (-1310 ± 21) km/s. This is best explained by the GW recoil *super-kick* scenario. This confirmation of the ejection process implies that the *final parsec problem* is resolved in nature, providing evidence that even the most massive black holes can merge.

**Main**

The powerful (bolometric luminosity [1] $L_{Bol}$ ~$10^{47}$ erg $s^{-1}$) radio-loud quasar (QSO) 3C 186 at a redshift z=1.068 lies in a massive (~$10^{11}$ $M_{Sun}$) galaxy located in the central regions of a cluster of galaxies [2]. The compact radio structure shows a young (~$10^5$ years old, [3]) one-sided jet with two hotspots on opposite sides of the radio core, and an S-shaped morphology [4]. This object was originally proposed as a gravitational wave (GW) recoiling black hole candidate based on SDSS optical spectroscopy and HST imaging and UV spectroscopy [1]. Theory predicts that such a phenomenon can occur after a black hole merger, under specific conditions of the progenitor systems. Anisotropic emission of GW during the final stages of the merger would result in the ejection of the merged black hole [5,6]. The main observable features characterizing a GW recoiling black hole [7,8] are a spatial offset between the QSO with respect to the photocenter of the host galaxy, and a velocity offset between the narrow and broad emission line systems. Being located at a distance generally smaller than 1 pc from the SMBH, the ionized gas in the broad line region (BLR) is gravitationally bound to it, while the narrow line region (NLR) is in the interstellar medium (ISM) of the host galaxy, on scales sometimes larger than 10 kpc. For a moving (recoiling) BH the two emission line systems should thus appear offset with respect to each other. For 3C 186, initial results showed a ~1".3 spatial displacement of the QSO (corresponding to ~10 kpc at the redshift of the object, Fig. 1a,b) and tentative evidence for a velocity offset between the broad and narrow line systems [1].

A definitive confirmation of these offsets, and a unique interpretation of 3C 186 as a GW recoiling black hole would constitute direct evidence that the theoretical limitation known as the *final parsec problem* [9] can be overcome. This problem is predicted to predominantly affect mergers of the most massive SMBH ($M_{BH} \sim 10^8$ $M_{Sun}$ and above). The theoretical framework predicts that in these events, after the pair shrinks by losing energy via dynamical friction and three-body interaction [10], it may stall at a distance of about 1 pc (depending on the mass of the pair). At such distances, the mass enclosed within the binary's orbit is smaller than the mass of the pair, and energy losses via the above mentioned mechanisms are no longer efficient. Energy loss via gravitational wave emission becomes predominant at significantly smaller distances, and thus the predicted timescale to coalescence of these very massive SMBH may exceed the Hubble time [9]. Recent evidence for a nanohertz GW background [11], when interpreted as a result of GW emission from merging SMBH, suggests that this problem is overcome in nature, with several proposed mechanisms, in both gas-rich and gas poor non-spherically symmetric galaxies, that may facilitate coalescence [76,77,78,79]. But direct evidence of coalesced SMBH remains elusive. A confirmed GW recoiling SMBH would provide key evidence. Several candidates have been proposed [12,13,14,15,16] yet none offer a unique interpretation, and prominent cases were later rejected [17, 18].

Several alternative scenarios for 3C 186 were discussed (and disfavoured) in previous work [1]: in contrast with the available observations, an eccentric disk emitter [19] and peculiar winds [20] would produce low and high ionization lines with different profiles, different apparent offsets, and significant variability; a double AGN was also considered unlikely based on both the absence of a second set of narrow lines [21], and the fact that the measured [OIII] luminosity matches the power of the observed QSO as a ionizing source. The interpretation of the QSO as a foreground source was firmly ruled out because of the detection of narrow C IV lines at the same redshift as the narrow emission lines, implying that the QSO, though moving, is currently located within the host galaxy. 3C 186 was later studied with a variety of instruments to pursue a deeper understanding of its nature and to find any clues that may rule out its interpretation as a GW recoiling BH. Deep HST observations confirmed the presence of the spatial offset between the QSO and the host galaxy [22]. The same study found the host to be dominated by a mature ($t >$ 200 Myr) stellar population, rejecting the hypothesis of either a recent or ongoing galaxy merger to explain the apparent offset between the QSO and the host center. NOEMA observations at mm wavelengths, revealed a large amount of molecular gas ($M_{H2} \sim 8 \times 10^{10}$ $M_{Sun}$) located at the systemic redshift of the host and spatially surrounding the host galaxy photocenter. This result formally ruled out the presence of an additional massive galaxy associated with the spatially offset QSO [23]. Furthermore, numerical relativity models were employed to assess the viability of a theoretical description of the object based on the initial results, in the framework of the GW recoiling BH scenario [24,25], and these models supported such an interpretation.

However, the low signal-to-noise ratio of the available spectroscopic data [1] precluded a robust and comprehensive interpretation, formally leaving alternative interpretations viable. The large errors in the velocity measurements (exceeding 20%) contributed to the uncertainty. Additionally, the unusual broad absorption features included in the spectral model introduced further ambiguity. Both the low S/N of the spectra analyzed in [1], and the fact that most of the emission lines in that work are blended with others contributed to such an ambiguity, making the spectral model less constrained. Therefore, the presence (and origin) of such features could not be firmly established. This work specifically addresses these key points, providing the decisive evidence that the modeled spectral features are real, and strengthening the interpretation as a GW recoiling SMBH.

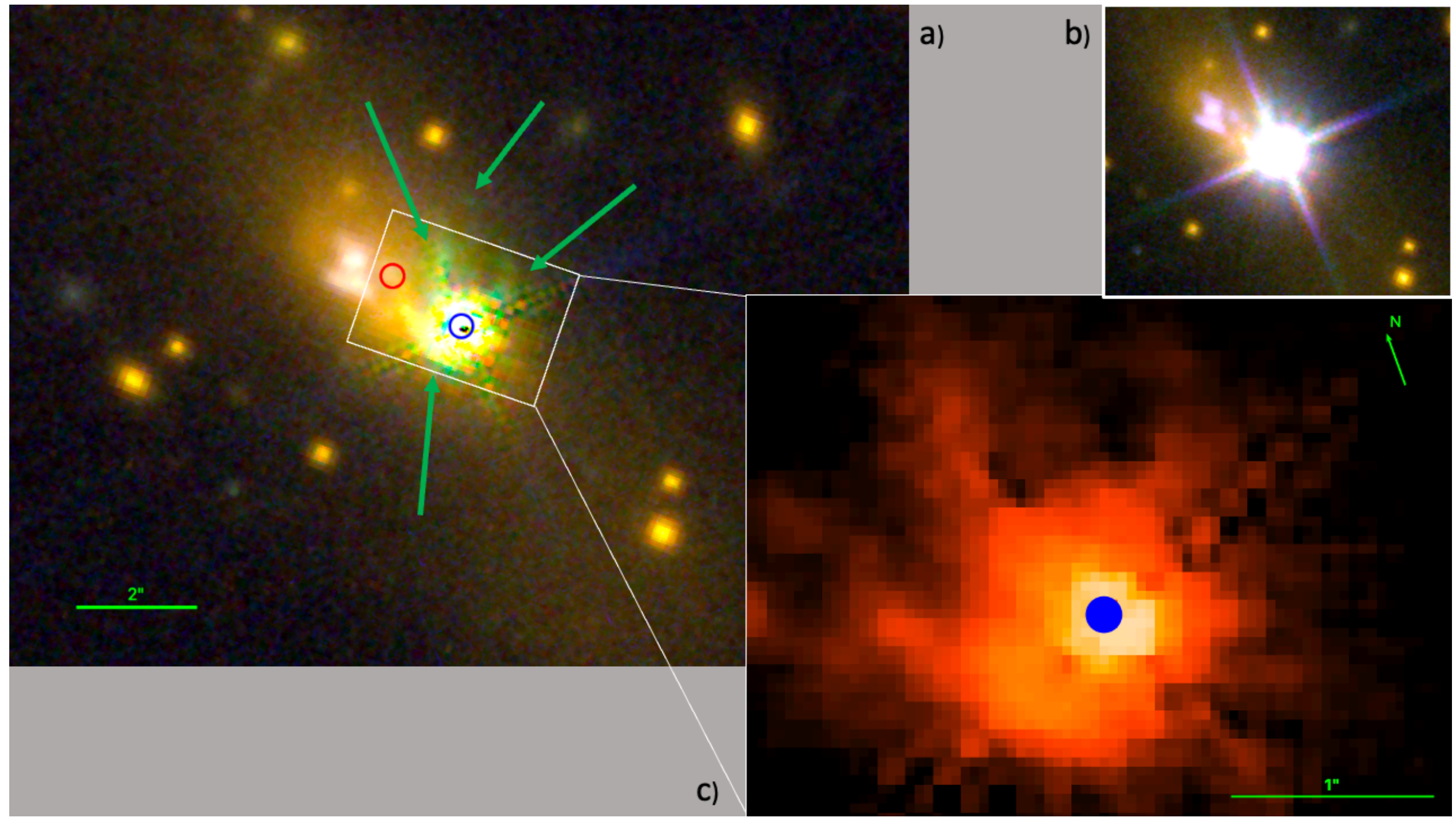

**Figure 1. The central region of the host galaxy, the QSO and its narrow line region.** The color image **(a)** shows an area of 15" x 11", corresponding to 121 x 90 $kpc^2$. North is up, East is left. The RGB image was obtained by combining three different filters from archival HST observations. The blue channel is ACS/WFC/F606W, the green channel WFC3-IR/F110W, and the red channel WFC3-IR/F160W. The red circle indicates the photocenter of the host as derived in [22], at R.A. = 7:44:17.586, Dec = +37:53:18.165. The location of the QSO (i.e. the black hole) is marked with a blue circle. The QSO's bright point source (point spread function, PSF) was subtracted off [19] and some residuals are visible in the core of the PSF and on the diffraction spikes. The features indicated by the green arrows just to the N and to the S-SE of the QSO are visible only in the F110W image (thus they are green in the RGB image). These are likely produced by [OIII] line emission in the narrow line region. The features to the North of the QSO are reminiscent of some of the biconical emission line regions often seen in AGN [26,27]. In **(b)** we show a region of 10"x10" centered on the QSO, before point-source subtraction. The bright point source is where the QSO power-law continuum and the broad lines seen in the spectra presented in this work are produced. Image **(c)** is a zoomed-in area around the QSO, as seen in the F110W filter. Here the host galaxy stellar emission was subtracted off to highlight the emission line gas morphology in the extended narrow line region. A region of r = 2 pixels (corresponding to 0".092) centered on the QSO is masked out in this figure.

## Results

### Spectroscopic evidence for a velocity offset in 3C 186

Here we present new spectroscopic observations of 3C 186 obtained with Subaru/SWIMS and VLT/XSHOOTER. All of the broad (defined as having FWHM > 3000 km/s) lines from permitted transitions included in the spectra (Hβ, Hγ, Mg II λ2800) appear asymmetric, with a concave blue side and a Gaussian shaped red side. All of these lines can be optimally fitted using a broad (FWHM ~ 5000 km/s) blueshifted absorption component in addition to the standard Gaussian broad emission line. Similar features are often observed in QSOs, the most extreme example being the broad absorption line quasars (BAL QSO [28]). Absorption lines in BAL QSOs have been detected for a number of lines, including CIV, MgII and, in a few cases, the Balmer series (see e.g. [80,81,82]). Broad absorption lines were also shown to be variable in some objects [83]. These features could be produced by partially ionized gas in a fast outflow located within the black hole sphere of influence, in the outer regions of the broad emission line region. Here we aim at determining the redshift of the BLR and compare it to the systemic redshift of the host galaxy. Therefore, it is crucial to separate the broad spectral component seen in emission (i.e. produced in the BLR) from the outflow seen in absorption, for each line in which both components are present.

Note that 3C 186 does not qualify as a BAL QSO only because the strict definition of such a class requires that the absorption of the continuum emission be greater than 10% (for a line of FWHM > 2000 km/s), which is not the case for the source studied here. However, 3C 186 could well represent an intermediate object between non-BAL QSO and BAL QSO. We tested that increasing the absorption by only a factor of 2 would be sufficient for the Hβ broad absorption component to qualify the source as a BAL QSO. The only exception is the broad semi-forbidden C III]λ1908 line, for which no significant absorption is neither needed in the best-fit model nor is expected. This line is collisionally excited, it has low oscillator strength, and therefore low probability for absorbing photons. The observed line in the spectrum of 3C 186 is in fact well fitted with a single Gaussian component.

All of the broad emission lines, produced in the QSO point-source within a pc from the SMBH (as seen in archival HST images, Fig. 1b), show a consistent offset with respect to the systemic redshift of the source, which is measured from the low-ionization narrow [O II]λ3727 line ($z_s$ = 1.06840 ± 0.00002, see Methods). The narrow emission lines are produced on larger scales within the galaxy ISM (Fig. 1c). A confirmation of such a velocity offset, in addition to the already well established spatial offset between the galaxy photocenter and the location of the QSO, implies that the best explanation is provided by a GW recoil super-kick causing the SMBH ejection after a merger of two SMBH of similar mass [5,6]. Given the large mass of the object ($M_{BH} > 10^9$ $M_{Sun}$), this provides the first direct evidence that the final parsec problem [9] can be overcome in nature even for the most massive SMBH. In the following, we use the terms blue-shifted and red-shifted with respect to the systemic velocity of the host, indicated by $z_s$.

In Fig. 2 we show the spectral region of the Hβ line taken with Subaru/SWIMS. This region also includes the [O III]λλ4959,5007 doublet. The [O III] lines are produced in the narrow line region, at large distances from the black hole (~ kpc scales) in the host galaxy ISM. The HST image shows that NLR is indeed extended (Fig. 1c). The flux measured from the narrow components of the emission lines included in the WFC3-IR/F110W filter bandpass as observed in the VLT spectrum (Hβ, [OIII]λλ4959,5007) matches the observed flux from the NLR in the HST image (see Material and Methods). This confirms both the extended origin of the NLR in this object, and the fact that the NLR is spatially centered on the QSO, ruling out the possibility of a dual AGN. The broad Hβ line is clearly asymmetric and best fitted with a very broad (FWHM = 10501 ± 333 km/s) blue-shifted (Δv = -1635 ± 253 km/s) component, and a blue-shifted (Δv = -4881 ± 131 km/s) broad (FWHM = 5916 ± 381 km/s) absorption component.

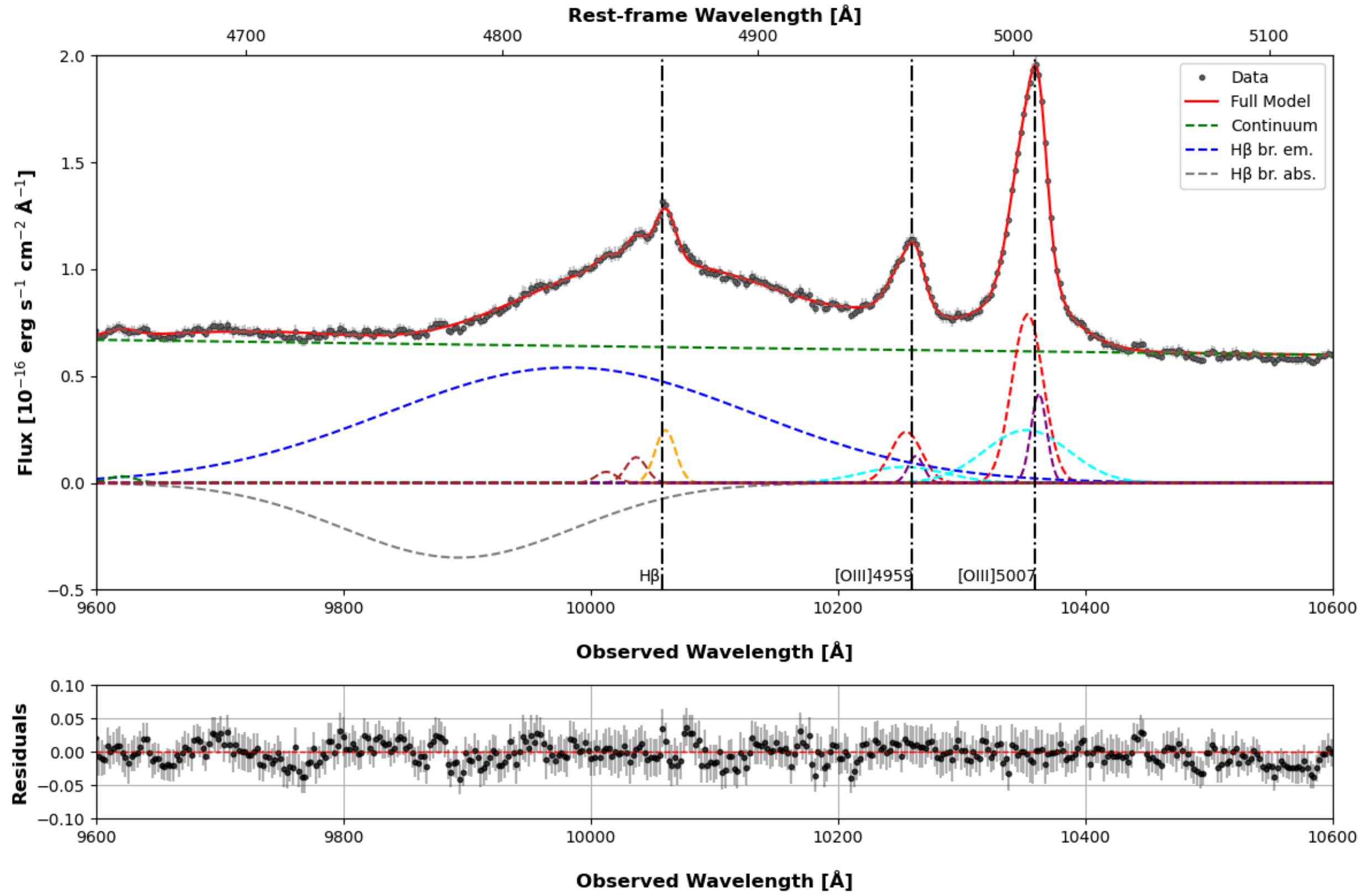

**Figure 2. Subaru/SWIMS spectrum of the Hβ + [OIII] line complex.**
The bottom x axis of the upper panel shows observed wavelengths and the top x axis shows the rest-frame wavelengths calculated assuming a redshift of $z_s$ (see text). Data points are black dots. Errorbars are at 2σ. The best fit model is shown with a red line on top of the data. Model components are shown as dashed lines: Hβ broad emission and absorption (blue and gray, respectively), [OIII]λλ5007,4959 blueshifted, redshifted and broad-ish (FWHM ~ 2500 km/s) components (purple, red and cyan, respectively, components a, b, and c in Table 1). The narrow components for Hβ are shown in orange (redshifted) and brown (two blueshifted components). The green dashed line is the power-law continuum. Residuals are shown in the bottom panel. Vertical **dot-dashed** lines indicate the wavelengths of the relevant emission lines calculated for the systemic redshift $z_s = 1.0684$.

In the Methods section we discuss all of the lines included in both the Subaru/SWIMS (Tab. 1) and VLT/XSHOOTER (Tab. 2) spectra in detail.

A new key piece of information is derived from the Mg II line profile, which shows two narrow spectral features at the line peak, superimposed to the broad components that Mg II shares with all other permitted lines (Fig. 3). A double peaked feature in which the bluer peak is brighter than the redder peak, is sometimes observed in some of the low ionization lines in QSOs and broad line radio galaxies, most notably in Balmer lines and Mg II [29,30,31,32]. This is routinely interpreted as produced in the outer regions of the accretion disk [33,34]. We modeled the line using a *KERRDISK* accretion disk model (Fig. 3) and derived that the emission is produced in a ring with $r_{in}$ = 820 ± 20 $r_g$ and $r_{out}$ ~ 2000 $r_g$, where $r_g = GM/c^2$ is the gravitational radius. The disk is blueshifted by Δv = -1294 (-25, +30) km/s with respect to the systemic redshift $z_s$. The disk inclination is not well constrained and we derived an upper limit at θ < 9 deg. The small value of the fitted disk inclination with respect to the line of sight is due to the small asymmetry of the double peak. Such a small angle is consistent with the fact that the radio jet is highly asymmetric (jet to counter jet ratio > 6, [4]) likely due to doppler boosting [35], implying that it is pointing at a small angle with respect to the line of sight. The best fit obtained with the disk model also includes blueshifted broad emission and a broad absorption line. The blue shift displayed by the Mg II broad emission line is consistent with that of the disk within 1σ. This consistency provides a key independent confirmation of the QSO velocity offset: the gas in both the BLR and the accretion disk are blueshifted, sharing the same velocity with respect to the host galaxy.

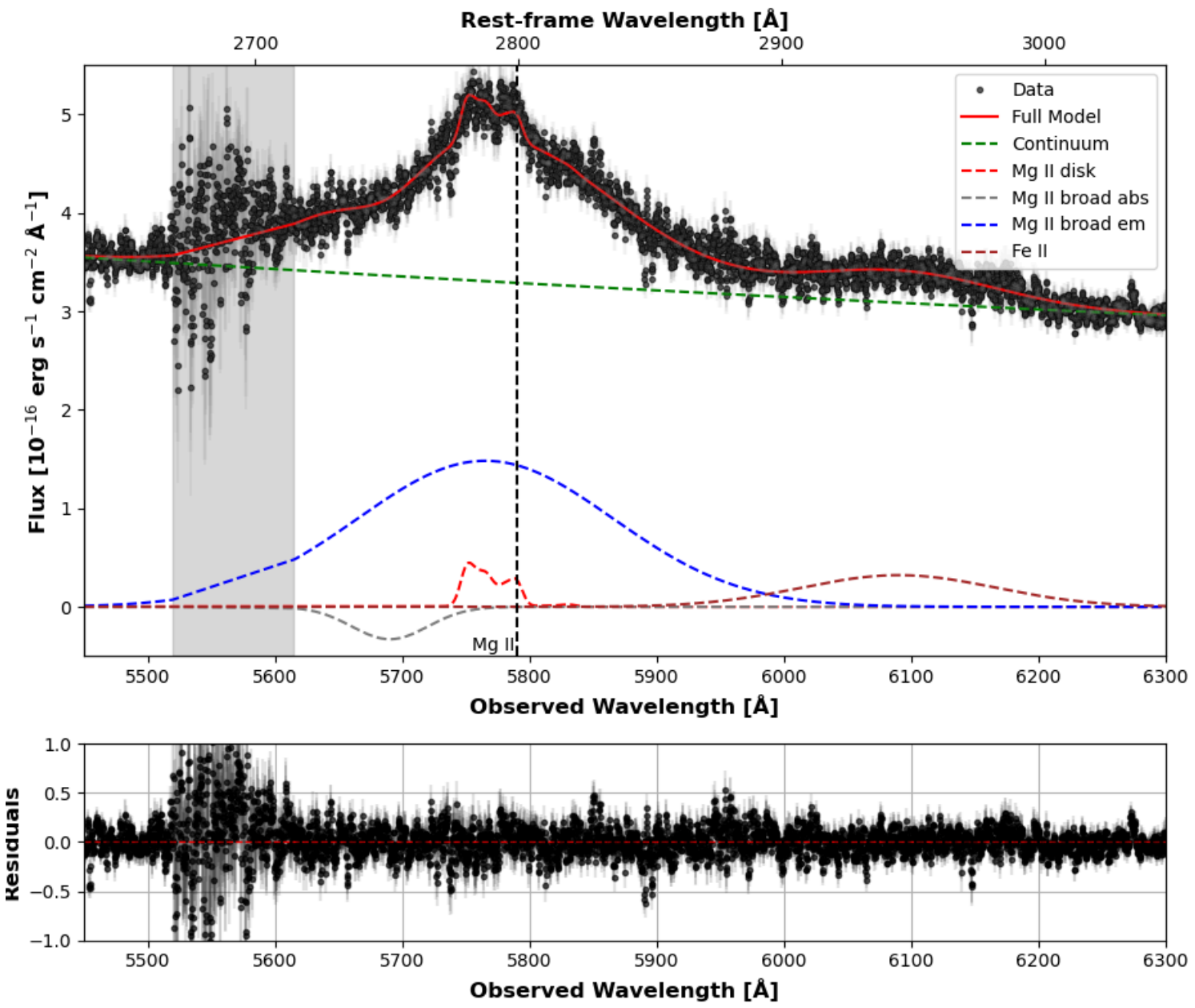

**Figure 3. VLT-XSHOOTER spectrum of the region of the Mg II 2800 line showing the double-peaked emission line produced in the accretion disk.**
The spectral region of Mg II as seen in the spectrum taken with VLT/XSHOOTER. The bottom x axis shows observed wavelengths and the top x axis shows the rest-frame wavelengths calculated assuming a redshift of $z_s$ = 1.0684. Data points are black dots. Errorbars are at 2σ. The spectrum was rebinned using a boxcar function with bin size = 5. Errors were summed in quadrature. The best fit model is shown with a red line on top of the data. Model components are shown as dashed lines: Mg II broad emission (blue), Mg II broad absorption (gray), Fe II emission (brown), *KERRDISK* model for the double-peaked accretion disk line (red). The continuum power-law is the green dashed line. Residuals are shown in the bottom panel. The shaded gray area corresponds to a spectral region with poor S/N due to the reduced sensitivity of the instrument's UVB and VIS spectrographs at the edge of their sensitivity range. The shaded gray area was masked out during the fitting process. The vertical dashed line indicates the wavelengths of the Mg II emission calculated for the systemic redshift $z_s$.

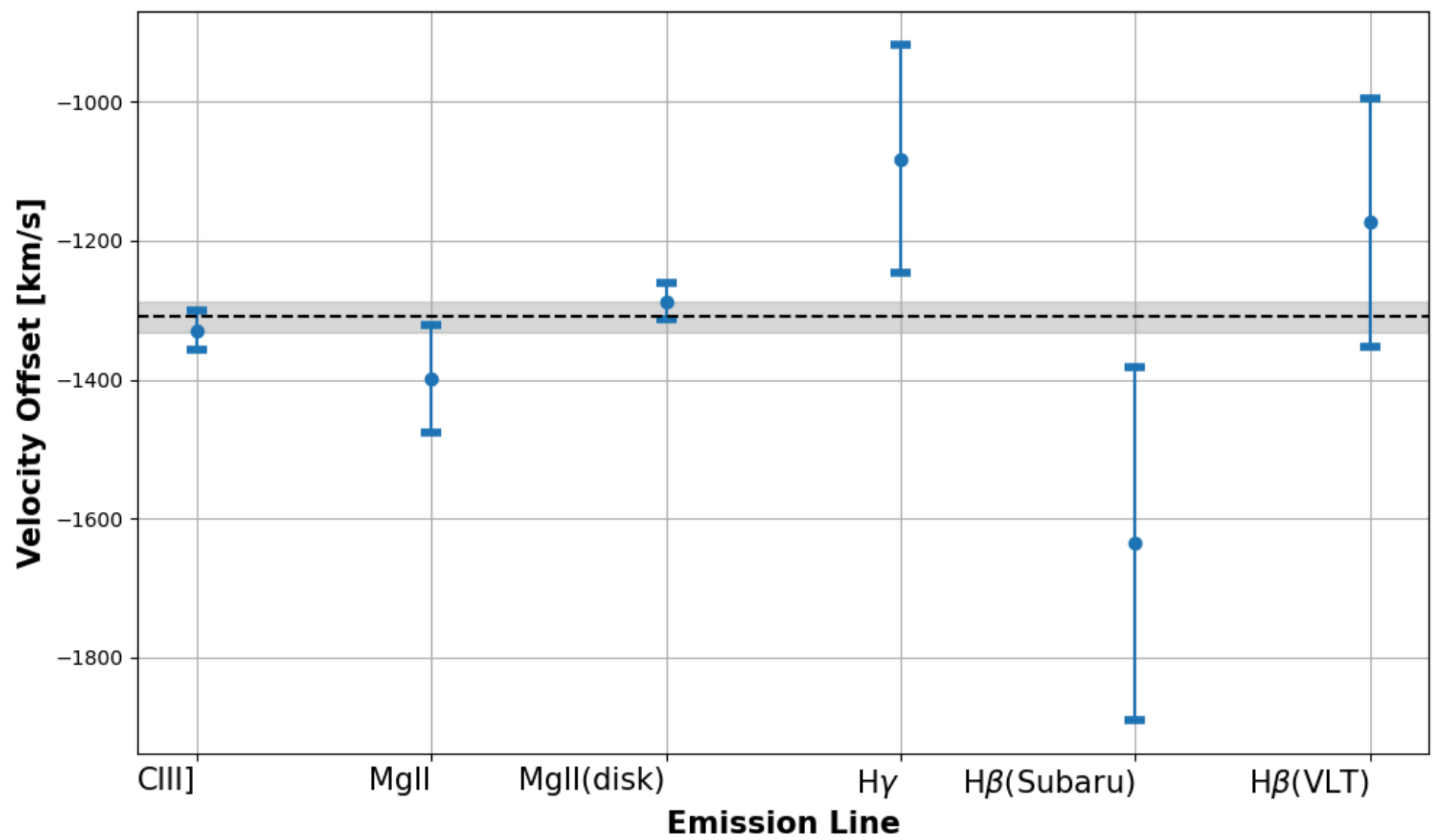

**Figure 4. Velocity offsets measured from the broad emission lines and accretion disk.** The figure shows all velocity offsets derived from each of the broad emission lines in the VLT/XSHOOTER and Subaru/SWIMS spectra discussed here, as well as the accretion disk double-peaked feature observed in the Mg II line. Each line is marked on the bottom axis and the corresponding velocity offset, calculated with respect to the systemic redshift of the host galaxy, is plotted with 1σ errorbars. The velocities measured from all of the spectral features are consistent within less than 2σ. The dashed horizontal line represents the weighted average of the values. The shaded area represents the reduced Chi-squared adjusted error on the mean.

In Methods we discuss the KERRDISK model in more detail. For consistency with the approach used for other spectral lines, we also present a simpler alternative model using two narrow (FWHM ~ 1000 km/s) Gaussian components to reproduce the double-peaked feature. Note that neither Gaussian is centered at the systemic redshift identified by all other narrow lines. While both models are statistically viable, we strongly favor the disk model because it provides a straightforward physical interpretation and yields results consistent with other observational constraints, as described above (disk orientation, broad line offset). The interpretation of the alternative Gaussian model faces significant physical challenges: the narrow line widths preclude a BLR origin, yet no other detected narrow emission lines show components at these velocities (see Table 2). This would require invoking an otherwise undetected narrow-line outflow that uniquely produces only Mg II emission at these specific blueshifted velocities.

We note that the double-peaked structure is clearly detected in Mg II but not in Hβ. This is not completely unexpected, as different emission lines may originate from different radii in the accretion disk or in the BLR, and are often found to be variable both in flux and in the observed profile [84]. There are known cases in which double peaked Mg II lines are present, while Hβ is either not double peaked, or it shows a very different profile (e.g. [85]). We do identify several faint narrow components in Hβ, including one blueshifted by ~500 km/s (Table 1, e.g. Hβ narrow em. b) that could potentially correspond to the red peak of the Mg II disk emission, though we refrain from overinterpreting these weaker features. Follow-up studies with simultaneous monitoring of both lines may shed further light on this issue.

By combining the offsets measured from all broad lines and the accretion disk features presented here, we derive an average velocity offset of ($\Delta v$ = -1310 ± 21) km/s. The offsets for each spectral feature and the weighted mean value are summarized in Fig. 4. In the figure we do not include the previously published measurements [1], since the error on the average velocity offset derived from those data is significantly larger (± 400 km/s) due to the lower S/N of the spectra. However, the results we present here are consistent with previous work within ~2σ.

In addition to measuring accurate velocity offsets for the BLR and the accretion disk, the spectra presented here allow us to derive important quantities for this QSO. Most importantly, using our measurement of the Hβ line FWHM we derive a BH mass estimate of **2.4** x $10^9$ $M_{Sun}$ (see Methods).

## Discussion

The data presented here provide decisive evidence that both the BLR and the accretion disk are blueshifted by the same amount with respect to the systemic redshift of the NLR and the host galaxy by ($\Delta v$ = -1310 ± 21) km/s. Both the accretion disk and the BLR lie within the sphere of influence of the SMBH ($r_{infl}$ ~ a few 100 pc [36]), and they are gravitationally bound to it. The size of the BLR is estimated to be 0.15 pc (see Methods), thus significantly smaller than the sphere of influence of the BH. The presence of such a significant velocity offset coupled with the already established spatial offset [1,22] is direct evidence that the SMBH (and anything that is bound to it) is being ejected from the host. An ejection due to three-body interaction would most likely leave the more massive black holes at the center of the host [74,75], with most of the surrounding gas remaining bound to the pair. Importantly, simulations show that for large SMBH masses an ejection is highly unlikely, since mergers of the triplet are strongly favored [74]. Furthermore, the measured BH mass of **2.4** x $10^9$ $M_{Sun}$ is aligned with the expectations from the standard galaxy magnitude-BH mass correlation [1], implying that the presence of two additional more massive black holes in the same system is highly improbable. A further, possibly even less likely scenario is that the broad line offsets are due to a supermassive black hole binary (SBHB). However, in that case the spatial offset between the host center and the SBHB would remain unexplained. These systems are believed to exist in the early stages of a galaxy merger, and observationally, all of the well established SBHB's with significant offsets are found in galaxies that are currently undergoing a merger, and with significant ongoing star formation [88]. 3C 186 would represent an unprecedented case of a massive offset binary in a mature, non-merging galaxy. While these alternative scenarios cannot be robustly ruled out, they would require additional ad-hoc assumptions to account for the observations. As already mentioned above, other scenarios such as an eccentric disk emitter, peculiar winds, chance superposition of a foreground source, were systematically considered and ruled out in [1] and [22].

The most plausible explanation that naturally accounts for all of the observed properties of 3C 186 is that the SMBH experienced a recoil kick following a merger of two SMBHs [5,6]. We stress that the presence of the broad (blue-shifted) absorption features in the permitted lines is unrelated to the interpretation of the object as a recoiling SMBH. As mentioned above, broad absorption lines are present in several (BAL) QSOs that are not recoiling, and likely originate in a fast outflowing wind just outside the broad emission line region, but still within the black hole sphere of influence. The offsets that point to the interpretation as a recoiling BH are measured between the broad emission lines and the narrow emission lines. Without the inclusion of a broad absorption component in all the permitted lines, the spectral model would require extra broad emission components that cannot be easily explained, but velocity offsets between narrow and broad lines would still remain (see Methods for more details on an alternative model for Hβ).

We utilize a numerical relativity model [25] to derive basic parameters for the GW event (see Methods for details). The main input parameters are given by the measured velocity of the QSO and the observed geometry [22]. Results show that the putative progenitor binary has a mass ratio $q = 0.56_{-0.31}^{+0.39}$. The spin of the more (less) massive BH is $\chi_1 = 0.84_{-0.54}^{+0.15}$ ($\chi_2 = 0.62_{-0.56}^{+0.35}$) and has an angle $\theta_1 = 92.6_{-49.3}^{+48.7}$ deg ($\theta_2 = 89.3_{-60.7}^{+65.6}$ deg) with respect to the orbital angular momentum of the binary. This is consistent with the so-called *super-kick* scenario [37,38], where the velocity of the ejected SMBH can reach up to 4000 km/s, or possibly more. Strikingly, while it was derived with a completely independent method, the angle between the spin axis of the BH and the observer's line of sight $\theta_{jet} = 8.6_{-6.4}^{+11.0}$ deg is fully consistent with the value of the disk orientation $\theta < 9$ deg obtained from the *KERRDISK* model best fit value. Finally, the geometry of the system requires a nearly aligned configuration with respect to our line of sight to justify the observations. Overall, these results are consistent with the picture presented in previous work [25] and further support the recoil hypothesis with the kick dynamics in general relativity.

The timescale from the kick can be derived from the numerical relativity model, albeit with a large uncertainty ($t_{GW} = 5.5_{-3.1}^{+15.1} \times 10^7$ yr). After the GW kick, the gas present around the merged BH is shaped as a punctured disk. Such a structure was dragged away with the BH and needs about $t_{visc} \sim 6.8 \times 10^7$ yr to fill the central gap and settle in a disk. As this process is completed, the AGN may turn on and the radio jet is launched. The radio source is young (i.e. $t_{AGN,rad} \sim 10^5$ yr), as estimated from synchrotron radiative cooling arguments applied to the radio source [3]. A lower limit to the age of the AGN consistent with the radiative cooling value can also be derived from the apparent size of the NLR's photo ionized gas in the HST image shown in Fig. 1a,c ($t_{AGN,ion} > 5 \times 10^4$ yr). The estimated survival time scale of the accretion disk is $t_{disk} \sim 3.3 \times 10^7$ yr. While these values can only be taken as order of magnitude estimates, the value of $t_{visc}$ may account for the delay between the GW kick following the BH merger and the time when the AGN turned on. A detailed derivation of these timescales is given in Methods and a summary figure is reported in Figure 5.

**3C 186 TIMELINE**

| PHASE | EVENT | DETAILS | TIME -----> | NOW | |
|---|---|---|---|---|---|
| 1 | BH Merger | - GW emission<br>- GW kick | <-- 2.5e7 - 2.0e8 yr --> | | |
| 2 | Accretion disk forms | - Time to fill the gap in the punctured disk | 6.8e7 yr | | |
| 3 | AGN turns on | - AGN radiation from disk photoionizes the NLR<br>- Radio jet is launched | 1.0e5 yr | | |
| 4 | Lifetime of the accretion disk | | | | 3.3e7 yr |

**Figure 5. Timeline of the GW kick and AGN activity.** This figure shows the estimated times scales of the events describing the 3C 186 GW kick and subsequent activity. Our observing time is marked with the black vertical bar (NOW). The first phase is the BH merger event. The time from the merger ($t_{GW} = 5.5_{-3.1}^{+15.1}$ x $10^7$ yr) is estimated from the numerical relativity MCMC model. In the figure we report the full range spanned by the result of the model given the error. Phase 2: the punctured accretion disk slowly fills the gap in $t_{visc} \sim 6.8$ x $10^7$ yr. Phase 3: the AGN turns on and it has been on for $t_{AGN,rad} \sim 10^5$ yr. Phase 4: the accretion disk survives for $t_{disk} \sim 3.3$ x $10^7$ yr.

While the estimate of the time from the GW kick is extremely uncertain, the fact that the QSO appears displaced by ~10 kpc (projected distance) implies that the physical separation between the host's center of mass and the QSO is larger than that value. Assuming a kick velocity of ~1300 km/s (see Table 6), and assuming constant velocity throughout a period of 5 x $10^7$ years, the derived displacement is ~50 kpc. The observed emission line filaments surrounding the QSO must therefore exist on those scales. This is not entirely surprising, since the host galaxy (which is the BCG of a cluster) is very large and extends (in projection) far beyond that size. Furthermore, it is known that the NLR in powerful radio galaxies may extend out to scales larger than 50 kpc [86,87].

In conclusion, we presented spectra taken with VLT/XSHOOTER and Subaru/SWIMS that provide the first unambiguous validation for a GW recoiling black hole interpretation of the peculiar features displayed by the powerful radio-loud QSO 3C 186. The existence of at least one very massive ($M_{BH} > 10^9$ $M_{Sun}$) object that is explained by such a phenomenon demonstrates that the *final parsec problem* can be overcome in nature. This provides compelling evidence that even the most massive SMBH can coalesce on a timescale significantly shorter than a Hubble time following galaxy mergers, and can grow in mass through such processes. This further supports the interpretation of the low frequency GW background recently detected by pulsar timing arrays [12, 71,72, 73] as due to SMBH close binaries. GW emission from single SMBH merger events may be detected with future space missions such as LISA (for SMBH of lower mass than 3C 186), and with pulsar timing array experiments [11,39,40].

## Methods

## Observations and data reduction

### VLT

We observed 3C 186 with VLT and the XSHOOTER echelle spectrograph [41] mounted on VLT-UT3 to get a full near-UV through near-IR spectrum including rest-frame UV and optical lines. Observations were performed in service mode on March 6th, 2023 (Program 110.23WJ, PI Chiaberge). We utilized slit mode (1".3x11", 1".2x11", 1".2x11" for the UVB, VIS and NIR arms, respectively). Nodding along the slit mode (5" length) was used for background subtraction in the NIR, and each exposure was 750s. The total integration time on source was 3000s. The seeing during the observations varied between 0".56 and 0".62.

The data were reduced using the standard XSHOOTER pipeline [42] and retrieved from the archive. Telluric correction was performed using the ESO tool Molecfit [43]. The final reduced spectra were converted from *fits* format to *ascii* tables and merged into a single file. A few (~20) single bad values associated with bad or hot pixels were manually removed. The merged spectrum was then rebinned using a rolling boxcar function with box size = 5. Errors were calculated using standard error propagation. For reference, in the spectral region just blue-ward of the Hβ line at 9800 Å (observer's frame) the S/N per resolution element is ~15, after rebinning.

### Subaru

We observed 3C 186 with Simultaneous-color Wide-field Infrared Multi-object Spectrograph (SWIMS)[45,46] mounted on the Subaru telescope, on Nov 25th, 2022 for half a night (S22B020, PI Morishita). We use the MOS mode of SWIMS, which offers a FoV of 6.6 × 3.3 arcmin$^2$, and set the slit width to 0".5. With the R~1000 slit-spectroscopy mode, the wavelength ranges of 0.9-1.4 μm and 1.4-2.5μm were simultaneously covered. The observation was aimed at specifically obtaining a deep view of the near-IR part of the spectrum in the observer's frame, which most importantly includes the Hβ and [OIII]λλ4959,5007 line complex. The observing condition was clear throughout the night, and the average seeing was ~0".4. The ABBA dithering strategy was adopted, with each consisting of 300 sec. The total integration time on source was 12,000 sec. A standard star (HIP26934) was observed for the telluric correction and the ThAr lamp was used for the wavelength calibration.

The data were reduced using standard IRAF tasks. The 1-dimensional spectrum was extracted via a box-extraction with the size of 20 pixels. The spectrum of the standard star was extracted in the same way and applied to the 3C 186 spectrum. Errors in the 1-D spectrum were estimated based on the pixel-by-pixel standard deviation of off-centered background regions. Poisson noise from the target was added in quadrature. The resulting S/N is significantly higher than for the VLT/XSHOOTER spectrum. In the region around 9800 Å (observer's frame) the S/N per resolution element is ~50.

Hubble Space Telescope imaging

HST images were obtained as part of program GO 15254 (PI M. Chiaberge) using ACS/WFC and the F606W filter, and WFC3-IR with F110W and F160W. The only filter that is strongly contaminated by significant line emission (Hβ and [OIII]λλ5007,4959) is F110W. The other two filters are dominated by continuum emission. The bandpass of F606W does include the Mg II line but that is spatially most likely limited to the bright point source (point spread function, PSF) from the QSO.

The observations, data reduction and PSF subtraction were described in detail in a previous paper [22]. An RGB color image using these three filters (R=F160W, G=110W, B=F606W) is shown in Fig. 1a). Images were drizzled to a pixel scale of 0".045. The central region of the galaxy before the QSO's PSF was subtracted is shown in Fig. 1b. In addition to the above steps already described elsewhere, here we performed host galaxy subtraction to remove the stellar emission of the host galaxy from the (PSF subtracted) F110W image and highlight the line emission only present in that filter. The host galaxy model was derived using *Galfit* [47] and the WFC3-IR/F160W image after PSF subtraction. The model was then renormalized and subtracted from the F110W image, which was also previously PSF-subtracted as previously described [22]. In Fig. 1c we show the central portion of the model subtracted F110W image. The central region of r = 3 pixels surrounding the subtracted QSO's PSF is masked out. This is because we believe the PSF subtraction in the innermost region of the PSF core may be highly uncertain. Furthermore, the emission in those pixels is dominated by the QSO's BLR and continuum emission, while the goal here is to focus on the narrow emission line morphology.

**Spectral fitting**

Modeling of the spectra was performed using *Specfit* [48] in *PyRAF*. The procedure we utilized is the same as described in [1]. We use extreme care in minimizing the number of components in order not to overfit the data by possibly including non-physical features. The basic components needed to optimally fit all of the broad lines are as follows: 1. One broad emission line, 2. One or more narrow emission lines, when needed, 3. One broad absorption line (except for C III], see below), 4. A power-law continuum. The central wavelengths of all of these components were left free to vary within a reasonable range (~100-200 Å for the broad lines, and ~ 10-30 Å for the narrow lines). All of these components are distinctive features observed in QSO spectra. Broad absorption lines are not observed in all objects but they characterize the subclass known as BAL QSO. These features are likely produced in a fast outflowing and partially ionized wind in the vicinity of the black hole, just outside of the BLR.

The final parameters for each line are determined when convergence is achieved. As recommended in the help file for *Specfit*, we utilize a combination of the *Simplex* and *Marquardt* minimization algorithms. Since the continuum slope is not constant over the large range of wavelengths spanned by our datasets, we fit each of the line complexes separately, but using the same procedure for all, for consistency.

In order to determine how many components should likely be included in each line complex, we used the composite QSO spectrum obtained by [49] as a guideline. The power-law continuum is fitted locally for each line complex, by choosing a region that is relatively free of other emission line contamination (e.g. from the Fe pseudo-continuum). The power-law index is almost constant in the bluer part of the spectrum (blue-ward of Hγ). A significant change in spectral slope is apparent around 8500 Å in the observer's frame. In this paper we refer to blueshifts with a negative velocity, and redshifts with a positive velocity.

In this paper, we adopted a ΛCDM cosmology [50]. The luminosity distance at the redshift of 3C 186, z = 1.0684, is 7275.3 Mpc and the corresponding projected scale is 1" = 8.244 kpc.

**Analysis of rest-frame UV and optical emission lines**

In the following we discuss each of the line complexes separately, and we detail model components and assumptions to derive the best fit for all lines. All of the lines discussed here, except for Hβ + [O III] which is covered by both Subaru/SWIMS and VLT/XSHOOTER, were only covered by the latter. Note that the Hα + [N II] complex was not covered by either instrument because such a line complex is redshifted to a spectral region strongly affected by atmospheric absorption bands, around 1.36 μm. Figure 6 includes all modeled lines and the best fit parameters are given in Table 2.

[OII]3727

A single Gaussian component is used to fit the [OII]λ3728 line. The derived line center is at λ = 7711.77 ± 0.06 Å in the observer's frame. This value is consistent with the result of [1] based on the SDSS spectrum. Using this line to derive the systemic redshift of the gas in the host galaxy and the vacuum wavelength for the strongest [OII] transition at 3728.38 Å (from the NIST database), we obtain z=1.06840 ± 0.00002. In Fig. 6a we show the spectral region of this line. The best fit model is the red line. The continuum is the green dashed line. The residuals show an apparent excess just redward of the center of the line. This is likely due to an outflow component that is also seen in other narrow emission lines, most prominently in [OIII]λ5007. However, for the purpose of deriving a reference value of the systemic redshift of the host we utilize a single gaussian for [OII].

C III]

This UV line is covered by the *XSHOOTER* UVB arm. We use two main gaussian components to fit this line: one narrow (FWHM ~1400 km/s) and one broad (FWHM ~ 8450 km/s). The broad component is clearly blueshifted with respect to both the systemic redshift (indicated by the vertical dashed line in Fig. 6b), and the narrow line. While the S/N is lower on the blue side of the line, one or possibly two more gaussian components in addition to the power-law continuum are needed to explain the excess emission observed at λ <~ 3850 and obtain a good fit. The origin of these components (both of which with FWHM ~ 9000 km/s) is uncertain. Tentatively, we identify the former with Si II (λ1816), and the latter with Al III (λ1857). If these are the correct lines they are both blueshifted by a similar amount as all other broad lines. However, since the identifications are uncertain and the S/N is extremely low we do not draw any conclusions based on this.

Mg II complex

The spectral coverages of *XSHOOTER*'s UVB and VIS arms overlap in the region just blue-ward of the redshifted Mg II line for our target, i.e. around 5500-5600 Å. Because of the low S/N at the blue end of the VIS wavelength coverage, we use the spectrum between 5450 and 6400, but we mask out wavelengths between 5520 and 5610 Å. We include in the first (rather simplistic) model the following components: one broad and two narrow (1 and 2 in Tab. 2) emission components for the Mg II line, one broad absorption Mg II line, and one additional broad emission line redward of the Mg II complex, likely due to Fe II emission that is often observed in bright quasars [49]. The model parameters are reported in Tab. 1 and the best fit is shown in Fig. 6c. The physical model using a Kerr accretion disk to produce the double peaked feature is discussed below.

Hγ complex

We consider the region between 8600 and 9250 Å to get a good handle on the continuum slope in this spectral region. Blue-ward of this region the Fe II features are strong, while large noise red-ward of 9250 Å is due to imperfect correction of significant telluric absorption. In addition to a prominent broad Hγ line, two narrow components are visible. One is likely Hγ from the NLR, and the other one is identified as the [OIII]λ4363 coronal line. In order to successfully reproduce the asymmetric shape of the broad line a broad absorption component must be included. The best fit is shown in Fig. 6d and the model parameters are reported in Tab. 1.

We note that the Hγ complex shows broad absorption that significantly affects the blue wing of the emission line, in line with the other broad lines discussed here. All broad absorption line offset are in the range spanned by BAL QSOs (i.e. -25,000 to 0 km/s, [91). The Hγ absorption shows a smaller velocity offset than Hβ (-2800 vs -4200 km/s) and has a broader FWHM (7500 vs 6500 km/s, see below and Tab. 2), potentially indicating stratification in the absorbing outflowing gas where Hγ absorption is produced at a different location, possibly closer to the black hole. However, we note that the detailed parameters of broad absorption features are particularly sensitive to continuum modeling choices, and while the asymmetric line profiles robustly establish the presence of absorption, some degeneracy exists in the precise absorption parameters.

Hβ + [OIII] complex

We restrict our model to a spectral region between 9600 and 10600 Å in the observer's frame. For wavelengths below this region the continuum is heavily contaminated by a number of blended Fe II lines. In addition, the S/N is lower in such a region because of the reduced sensitivity close to the boundary of the instrument coverage towards lower wavelengths. This holds both for the VLT/*XSHOOTER* and the Subaru/SWIMS spectra.

We first focus on the Subaru spectrum, which has the highest S/N. The most prominent lines in this region are the [OIII]λλ5007,4959 doublet and the broad Hβ line. Hβ appears highly asymmetric and with a convex blue side. Two (or possibly three) small peaks near the peak of the broad line are also visible, slightly above the noise level. We fit those features with three (narrow) gaussian components. The concave shape of the blue side of the Hβ line is well modeled with the inclusion of a broad absorption line which appears significantly blue-shifted. The model with the above six components is sufficient to describe the complex, but the [OIII] doublet shows significant residuals in a blue wing for both lines. In order to optimize the fit, and bearing in mind the detailed results of Keck/OSIRIS IFU observations which show significant velocity structures for [OIII]λ5007 with both blue and red shifted components [51], we include three Gaussians for each [OIII] line. The first ([OIII],a in Table 1) is slightly redshifted, the second ([OIII],b in Table 1) is consistent with being in the rest frame, or slightly blueshifted, and the third ([OIII],c in Table 1) is broad-ish (FWHM ~ 2400 km/s) and significantly blueshifted. We fix the ratio between each respective component of the [OIII] doublet to the expected value of 3 to 1. Contribution from a broad Fe II line may also be present in QSOs. Such a component is often observed between [OIII]λ4959 and Hβ, at around 4930 Å [49] but in the case of Population B quasars such a component is very faint [52]. We tried to include this component in the model but the results are unconstrained and the fit does not improve. We conclude that there is no evidence that Fe II contributes to this spectral region. The small residual around 9700 Å is likely due to a faint broad He II emission. We tested that the inclusion of such a line in the model does not change the measurements for the brighter lines in the spectrum, and most importantly it does not impact the value of the blueshift for the broad Hβ. The best fit is shown in Fig. 2 and the model parameters are reported in Tab. 1.

The S/N of the VLT/*XSHOOTER* spectrum is significantly lower but to check for consistency we fit the same range of wavelengths (9600-10800 Å) using the same model used for the Subaru data, with only one notable difference. Since multiple narrow line components for Hβ do not seem to be visible, thus we only include one narrow Hβ component in the model, in addition to the broad emission and absorption lines. The best fit is shown in Fig. 6e and results of the fit are in Tab. 2. The values of the velocity offset of the broad line with respect to the systemic redshift measured in the two spectra are consistent within 2σ. The estimated errors on the model parameters for the two spectra are similar, particularly for the broad components, which are bright, they extend over a wide range of wavelengths and are well determined even at a lower S/N. However, we note that for some of the narrow emission lines, the model parameters of the VLT spectrum are significantly more uncertain, likely as a result of the lower S/N.

Note that the derived fluxes for each component in the VLT spectrum are not consistent with the measurements derived from Subaru. While an accurate flux calibration is not the main goal of this work, we believe there could be a straightforward explanation for this. For both the continuum and the broad lines, some level of variability is expected, since the observations were not taken simultaneously. For the narrow lines, variability cannot be the origin of the discrepancy between VLT and Subaru. The difference between two observations of [OIII]λ5007 components a and b, where the flux in Subaru/SWIMS is lower than the one measured from VLT/*XSHOOTER*, is likely due to the fact that the SWIMS slit is significantly narrower than the one used with *XSHOOTER*, and the NLR extends well beyond the slit width, as seen in Fig. 7. Interestingly, component c (faster, broadish outflow) does not significantly differ between the two datasets, likely indicating that this component is more compact and more centered on the QSO. These components are reminiscent of the results obtained with Keck/OSIRIS IFU data, which showed evidence for strong outflows within ~1" of the QSO, traced by the [OIII]λ5007 emission line, with velocities ranging from $\Delta v \sim +100$ to $\Delta v \sim -700$ km/s. Some of these components do not spatially coincide with the QSO point-like continuum source [51].

The flux of the narrow lines measured from the VLT spectrum were used to spatially identify the NLR in the HST image. We co-added all of the narrow components in the best fit model to derive the total flux of the lines included in the F110W filter bandpass, i.e. Hβ (narrow) and all of the [OIII]λλ5007,4959 components. The total flux is $F_{line} = 2 \times 10^{-14}$ erg cm$^{-2}$ s$^{-1}$. This emission line flux, using *pysynphot* and the WFC3-IR exposure time calculator is predicted to generate ~177 e$^-$/s in the image. We performed aperture photometry in a circular region with r = 0".6 (i.e. diameter is the same as the XSHOOTER slit width) centered at the location of the QSO, masking out the central r=2 pixels (corresponding to 0".092) because of imperfect PSF subtraction in the innermost region of the PSF. The measured flux is 163 e$^-$/s, roughly consistent with the expected flux from *pysynphot*. We therefore conclude that the structure observed in Fig. 1c (and in Fig. 7) is the NLR of 3C 186.

**Accretion disk model for Mg II**

A line profile showing two narrow peaks superimposed to a broader Gaussian component are characteristic of double peaked broad line sources [30]. Since double peaked low ionization lines are routinely interpreted as the blue and red-shifted sides of an accretion disk, we utilized the *KERRDISK* relativistic disk model [53] in XSPEC 12.14.1 [54], convolved with a Gaussian smoothing to match the instrumental resolution. The method is the same as for the analysis of the spectrum of the Seyfert galaxy NGC 3147, which also shows similar features produced in the accretion disk [55, 56]. The line emissivity index was fixed at 3, and the BH spin is unconstrained, as expected since the inner radius of the emitting region turns out to be much larger than the innermost stable orbit. The disk inclination, inner and outer radii that better fit the line profile are reported in Table 3, together with the total flux of this line component and the corresponding velocity shift. Relevant best-fit model parameters are given in Table 3. Note that the best fit values of the broad emission and absorption components derived using the *KERRDISK* model are consistent with the results from *specfit*, within $\lesssim 2\sigma$.

A two component (disk + BLR) model has recently been found to be appropriate for the double peaked broad Hβ line in the QSO SDSS J125809.31 + 351943.0 [57]. The broad component for the BLR was found by those authors to be variable and sometimes dominates over the double peaked disk component in that source. At much lower AGN luminosities, the Seyfert galaxy NGC 3147 was also found to display a double peaked accretion disk feature in the Hα line, superimposed to a broad gaussian component from the BLR. Both components were found to be variable on time scales of years [56].

A similar case of a lower redshift (3C 47, z = 0.4) radio loud QSO is discussed in [58], where these authors use a relativistic accretion disk model for the double peaked Mg II emission. However, in 3C 47 the two peaks are centered at the systemic redshift of the source. In 3C 186 the disk line is significantly blueshifted ($\Delta v = -1288_{-25}^{+29}$ km/s) and the blueshift of the disk is consistent with that shown by the gaussian component of all broad lines (including that of Mg II).

Note that the absence of a clear double peaked line profile in other lines is not completely surprising. The line that is known to display the most prominent double peaked structure is Hα, which cannot be observed from the ground since it lies within one of the atmospheric absorption bands.

**Alternative model for Hβ**

We considered an alternative spectral model that does not include a broad absorption line. In order to fit the asymmetric shape of the broad line at the same level of accuracy as for the model with broad absorption, we need two broad emission components. The best-fit model accurately reproduces the observations but the second broad component at 10202 Å does not obviously correspond to any known emission line (Tab. 4, Fig. 8). Possible identifications could be 1. redshifted Fe II λ4924 (which corresponds to 10184 Å in the observer's frame); 2. a redshifted Hβ wind. Multiple broad emission components are often used to model spectra of powerful QSO and are usually attributed to the presence of winds [59]. In the former case the line would be significantly broader (FWHM = 7543 ± 438 km/s) than the typical Fe II emission at this wavelength observed in powerful quasars, which is on average a factor of >10 narrower than Hβ [49]. In the latter, the fact that such a wind component would be broader than the main Hβ line component, combined with the extreme wind velocity (Δv = 4316 km/s), makes such a framework very hard to physically interpret. Furthermore, we note that even if we accept this model as correct, regardless of its physical interpretation, the main component of the broad Hβ line would still be significantly blue shifted. However, two results would be in contrast with what we measure for all other broad lines. Firstly, the blue-shift velocity is only Δv = (-598 ± 57) km/s. While the presence of a blueshifted broad line still fits in the GW recoiling BH scenario, the velocity for this specific line is inconsistent with the blueshifts measured in all other lines. Secondly, the FWHM of the Hβ line as derived in this framework (6146 ± 138 km/s) is significantly smaller than that of all other broad lines considered in this work, in contrast with the typical behavior observed in large samples of QSOs [49]. Therefore, we conclude that the model that includes a broad absorption component is superior to a model with multiple broad emission components since the latter leads to inconsistent results.

Note that the asymmetric shape of other lines such as Mg II could also be reproduced using a model with two broad lines, similarly to what we showed for Hβ, and as accurately as for the models using broad absorption. We stress that the main inconsistency with the alternative model is related to the fact that if such a "double BLR" existed in reality, we should also see an additional red-shifted broad emission line in C III]. However, no emission features are seen red-ward of the C III] line. This fundamentally rejects a model with two broad line regions.

**Black hole mass, Eddington ratio and other physical quantities and relevant timescales**

We utilize the Hβ line from Subaru/SWIMS, which is our highest S/N information, to derive a measurement of the black hole mass in 3C 186. Using the formula in [60] for this line,

$$M_{BH,H\beta} = 10^{6.7} \times (FWHM_{H\beta} / 10^{3}\,\mathrm{km\,s^{-1}})^{2} \times (\lambda L_{\lambda(5100\,Å)} / 10^{44}\,\mathrm{erg\,s^{-1}})^{0.5}\,, \quad (1)$$

and using the FWHM and $L_{5100}$ as derived from the Subaru data presented here, we derive a BH mass of $M_{BH,H\beta} = 2.4 \times 10^9 M_{Sun}$, in line with previous estimates [1]. As a sanity check, we also use the Mg II line to derive an alternative estimate for the BH mass [60] and we obtain $M_{BH,MgII} = 4.7 \times 10^9 M_{Sun}$. The values of the continuum luminosity and line FWHM used in these calculations are reported in Table 5, A discrepancy of a factor of ~2 is in line with results from the literature [61,62]. Whenever needed in this paper we utilize the result derived from the Hβ line since it is often assumed to be slightly more accurate [63]. We stress that the accuracy of the BH mass estimate predominantly depends on the intrinsic scatter of the used relations, which is generally assumed to be a factor of ~ 2 - 3.

The Eddington ratio can be derived using the bolometric luminosity of the source $L_{bol} = 7.5 \times 10^{46}$ erg $s^{-1}$ estimated from the [OIII]λ5007 luminosity [1] and the black hole mass. We derive a value of $\varepsilon_{Edd} = L_{bol} / L_{Edd} = 0.17$, in the normal range for powerful quasars [64].

Following [62] we can estimate the radius of the BLR

$$R_{BLR(H\beta)} = 538.2 \times (L_{5100} / 10^{46} \text{ erg s}^{-1})^{0.65} \text{ light-days.} \quad (2)$$

We obtain $R_{BLR}$ = 176 light-days = $4.6 \times 10^{17}$ cm = 0.15 pc. The sphere of influence of the black hole, for a BH of ~ a few $10^9 M_{Sun}$, is $R_{infl}$ ~ a few hundred pc [36]. This holds if the SMBH were located at the center of the host, where the stellar density (and thus the stellar velocity dispersion used to estimate $R_{infl}$) is higher than at the current location of the offset SMBH in 3C 186. However, the outer radius of the region that is carried away by the recoiling BH is likely much smaller (< a few pc, or of the order of $R_{BLR(H\beta)}$).

The survival time of the accretion disk around an ejected SMBH can be estimated from [7]

$$t_{disk} \sim 8.4 \times 10^6 \alpha_{-1}^{-0.8} \eta^{0.4} M_7^{1.2} v_8^{-2.8} \text{ yr,} \quad (3)$$

where $\alpha_{-1}$ is the viscosity parameter of the disk in units of 0.1 (and assumed to be $\alpha_{-1} = 1$), $\eta = (\varepsilon / 0.1)/(L_{bol} / L_{Edd})$, ε being the radiative efficiency of the disk, $M_7$ is the BH mass in units of $10^7 M_{Sun}$ and $v_8$ is the BH kick velocity in units of $10^8$ cm $s^{-1}$ = $10^3$ km $s^{-1}$. Assuming ε = 0.1, and $L_{bol} = 7.5 \times 10^{46}$ erg $s^{-1}$ [1], we obtain $t_{disk} = 3.3 \times 10^7$ yr. Note that the value of $t_{disk}$ was corrected by a factor of $(R_{max,3}/45\ v_8^{-2})^{1.4} = 0.0067$ because in our case the outer radius is larger than the maximum radius ($R_{max,3}$ in units of $10^3$ Schwarzschild radii, $R_S$) allowed for the disk not to exceed the mass of the BH itself (note (30) in [7]).

In the GW recoil scenario the accretion disk is formed by material that surrounded the BH pair before coalescence. Simulations [89, 90] show that as the pair gets closer, a gap opens across a region surrounding the pair and the disk becomes truncated (or punctured). The time to fill the gap after coalescence can be estimated using formulae (1) and (3) in [7]

$$t_{visc} (R_{in}) \sim 4.1 \times 10^4 \alpha_{-1}^{-0.8} \eta^{0.4} M_7^{1.2} R_{in,3}^{1.4} \text{ yr,} \quad (4)$$

where $R_{in,3} \sim 0.65\ \alpha_{-1}^{-0.31} \eta^{0.15} M_7^{0.077} = 1.22$ is the inner radius (in units of $10^3 R_S$) of the punctured disk for the parameters used above. We thus derive $t_{visc} = 6.8 \times 10^7$ yr.

The survival time of the accretion disk can also be derived by observing that $t_{disk} \sim R_{disk}/(\alpha\ c_s)$ where $c_s$ is the sound speed in the disk. The size of the disk in a recoiling BH can be assumed to be the size of the entire region carried away by the BH, i.e. the disk and the BLR, which is $R_{BLR(H\beta)} \sim 0.25$ pc, as derived above. α is set to 0.1 as above and $c_s$ is ~ 3 x $10^5$ cm $s^{-1}$ in the inner part of the disk, and drops down to ~$10^4$ cm/s at $R_{BLR}$. Thus $t_{disk} \sim 10^7$ yr, in agreement with the estimate given by (3).

Another relevant timescale can be inferred from the observed size of the narrow line region as seen in the HST image (Figs. 1c, 7). The apparent extension is ~2", which corresponds to ~16.5 kpc projected size, at the redshift of the source. This sets a lower limit to the AGN of $t_{AGN,ion} > 5 \times 10^4$ yr, which is consistent with the radiative age of the radio source of $t_{AGN,rad} \sim 10^5$ yr [3].

Note that these should be taken as order of magnitude estimates, or at best as accurate to within a factor of ~ a few, given the uncertainties in the underlying physical parameters (viscosity parameter α, disk structure, etc.).

Numerical relativity model for the gravitational wave black hole kick scenario

An early computation of the progenitor's parameters based on phenomenological formulae and the earliest available measurements from [1] was performed by [24]. In this work, we use a state-of-the-art numerical relativity surrogate model for remnant properties, in combination with updated velocity measurements from this work and the information on the projected kick and spin vectors derived in [22]. As a result, we are able to solve the full geometrical configuration of the post-merger SMBH. Moreover, the MCMC analysis allows us to qualitatively assess the likelihood of such a system within the full parameter space.

Assuming the recoil hypothesis, we use the statistical procedure presented by [25] to identify the parameters of the progenitor BH binary that are compatible with the new observations presented in this paper. First, we impose a constraint on the recoil velocity projected along the line of sight of $v_{\parallel} = -1310 \pm 21$ km/s (see Main). Second, we require the projections of the kick and the spin of the remnant BH onto the plane of the sky to be perpendicular to each other ($\theta_{v\chi,\perp} = 90° \pm 10°$; [22]). The 10° uncertainty is a conservative estimate given the number of significant figures reported by [22]; our earlier analysis [25] shows that this error has a minimal impact on the results.

We perform a Bayesian analysis using the emcee Monte Carlo Markov Chain (MCMC) sampler [65]. We choose a bivariate Gaussian likelihood centered on the aforementioned constraints and a standard deviation corresponding to measure uncertainties. We assumed a quasi-circular progenitor BH binary characterized by mass ratio $q = m_2/m_1$, where $m_1$ ($m_2$) identifies the mass of the heavier (lighter) BH. The dimensionless spin parameters are defined in terms of the magnitude $\chi_{1,2} \in [0,1)$, the polar angles $\theta_{1,2}$ with respect to the orbital angular momentum, and the azimuthal angles $\phi_{1,2}$, which are defined in the orbital plane. We model the difference $\Delta\phi=\phi_2-\phi_1$ because it better traces the occurrence of large recoils [66,67,68]. The prediction of the post-merger BH properties, namely the final spin $\boldsymbol{\chi}_f$ and the recoil velocity **v**, is performed using the machine-learning surrogate model NRSur7dq4Remnant by [69], which is trained on numerical-relativity simulations of quasi-circular merging BHs with mass ratios $q > 0.25$ and spin magnitudes $\chi_{1,2} < 0.8$. All vectorial quantities are specified in a reference frame where the *z* axis is aligned with the orbital angular momentum, the *x* axis extends from the lighter to the heavier BH, and the *y* axis completes the right-handed triad. All pre-merger quantities are defined at $t = -100G(m_1+m_2)/c^3$ prior to the merger.
We set uninformative priors on the binary parameters and allow the surrogate model to extrapolate predictions in its safe regime of validity [69] with $q > 1/6$ and $\chi_{1,2} < 1$. In particular, the mass ratio q and the dimensionless spin magnitudes are sampled uniformly in [1/6, 1] and [0,1), respectively, while the spin directions are isotropically distributed. Two additional isotropically distributed angles, $\theta_n$ and $\phi_n$, are employed to model the observer line of sight.

The results of the MCMC analysis are presented in this Section. The sampler converges to a region of the parameter space that is consistent with the recoiling hypothesis. Figure 9 illustrates the posterior distributions of the sampled parameters. As in our previous analysis, there are two possible configurations, where the remnant spin and the line of sight are either aligned ($0° \leq \theta_{n\chi} < 90°$) or anti-aligned ($90° \leq \theta_{n\chi} < 180°$). Compared to previous work [25], a lower constraint on the kick velocity projected along the line of sight corresponds to a smaller recoil. The resulting system appears asymmetric in the progenitor masses (although the uncertainty is significant since we obtain $q = 0.56_{-0.31}^{+0.39}$). Further details can be found in Table 6. We also build posteriors for the remnant properties; in particular, we find that the mass of the BH is $m_f = 0.96_{-0.02}^{+0.01}$ M (where M is the total mass of the progenitor binary), the spin magnitude is $\chi_f = 0.69_{-0.14}^{+0.14}$ , and the recoil velocity is $v = 1328_{-42}^{+61}$ km/s. In both configurations, we confirm that the recoil velocity is nearly aligned with the observer ($\theta_{nv} = 8.5_{-6.2}^{+9.9}$ deg). The spin and velocity components onto the plane of the sky are small and appear perpendicular due to a strong projection effect. Moreover, by assuming that the jet is launched along the direction of the BH spin, we infer a viewing angle, i.e. the angle between the observer and the AGN jet, $\theta_{jet} = \min(\theta_{n\chi}, 180°-\theta_{n\chi}) = 8.6_{-6.4}^{+11.0}$ deg. This result is consistent with the observation of the Mg II line from the accretion disk (see Main) and the jet-to-counterjet ratio [4].

Figure 10 presents a comparison of the prior and posterior distributions for the two quantities that directly enter the likelihood. This system still appears to be quite rare, with the two distributions exhibiting minimal overlap due to the fine-tuning required for large recoils. From this reconstruction of the system, one can also determine the elapsed time since the merger. The astrometric observations indicate a displacement of $d_\perp = 11.1 \pm 0.1$ kpc. Assuming a constant velocity after the coalescence, the Bayesian estimate for the merger time is $t = 5.6_{-3.1}^{+15.1} \times 10^7$ yr.

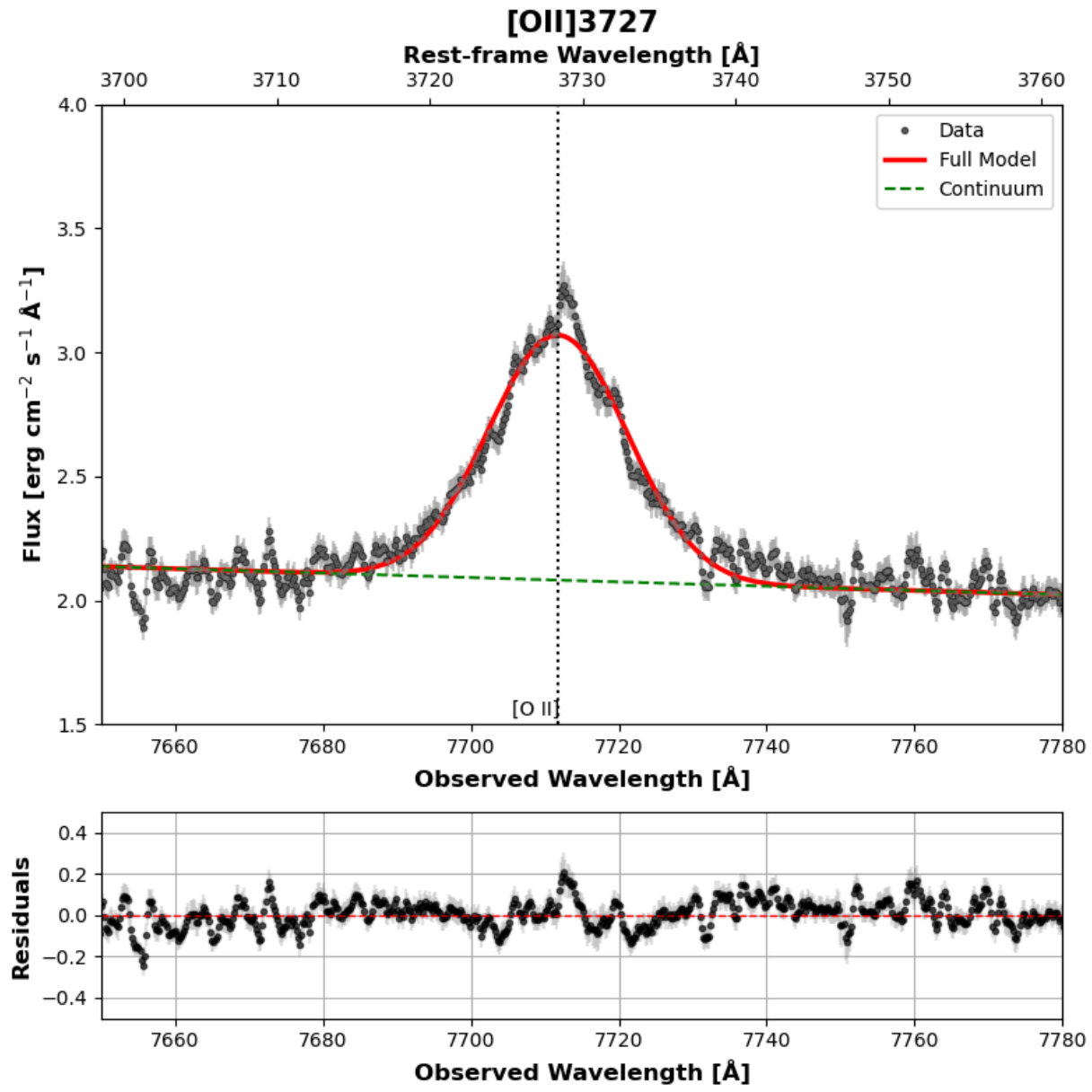

**Figure 6a VLT/XSHOOTER spectrum of the** [OII]λ3727 **emission line, with best fit model.** The figure shows the spectral regions around the [OII]λ3727 emission line from VLT/XSHOOTER. The data points are shown as black dots. Errobars are at 2σ. The x bottom axis reports observed wavelengths, while the top axis shows rest-frame wavelengths calculated using a systemic redshift $z_s$ = 1.0684, as derived from the [OII] line. The full model is shown in red, and the power-law continuum with a green dashed line. Gaussian model components are shown with dashed lines in different colors. The main model components are described in the inset. The bottom panels report the residuals. The vertical line indicates the rest frame central wavelength of the line.

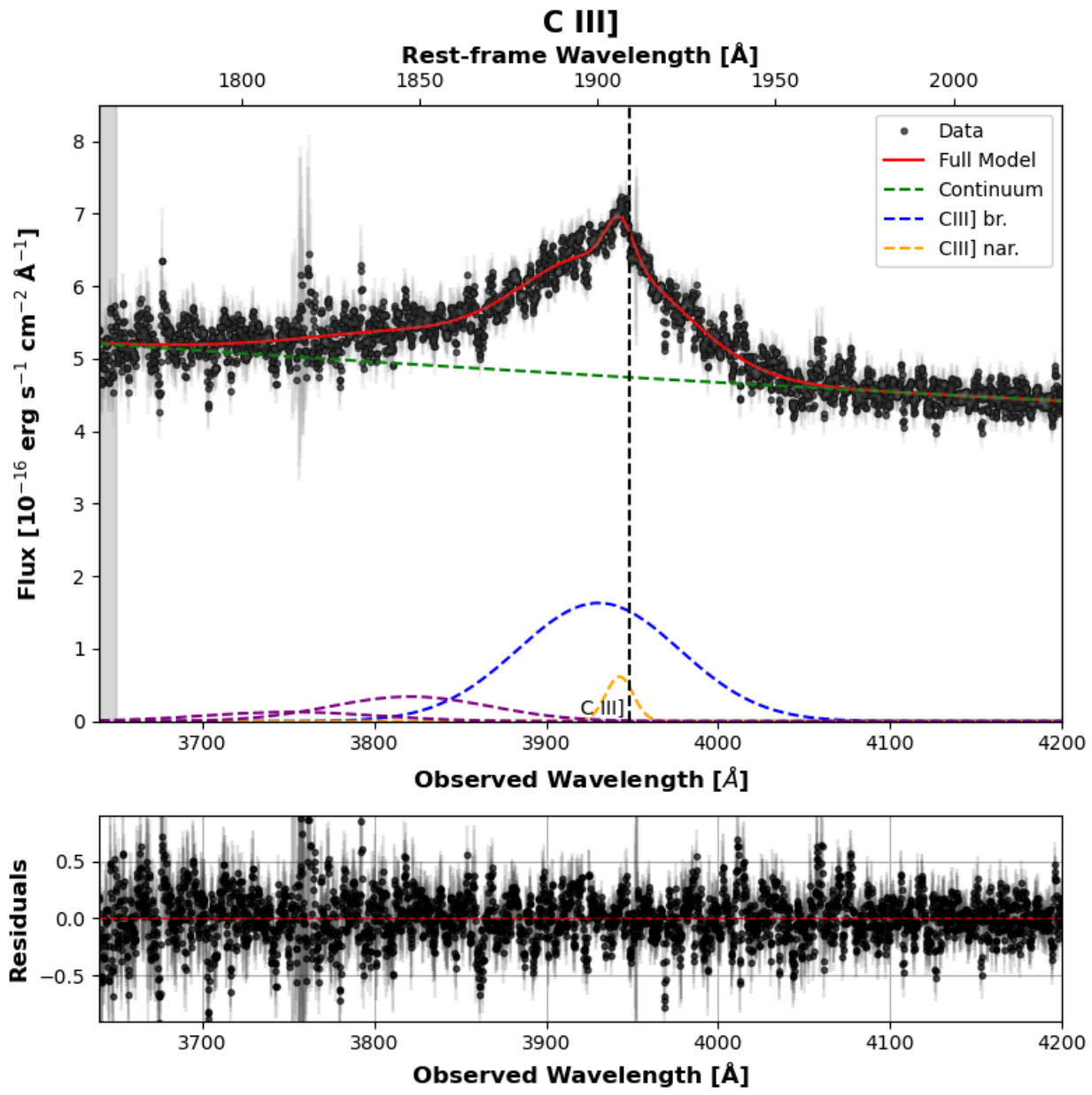

**Figure 6b. VLT/XSHOOTER spectrum of the** C III] λ1908 **emission line, with best fit model.** Same as for Fig 6a, but for C III] λ1908. Shaded gray areas indicate regions that were masked out during the fitting process because of enhanced noise. The vertical line indicate the rest frame central wavelength of the line.

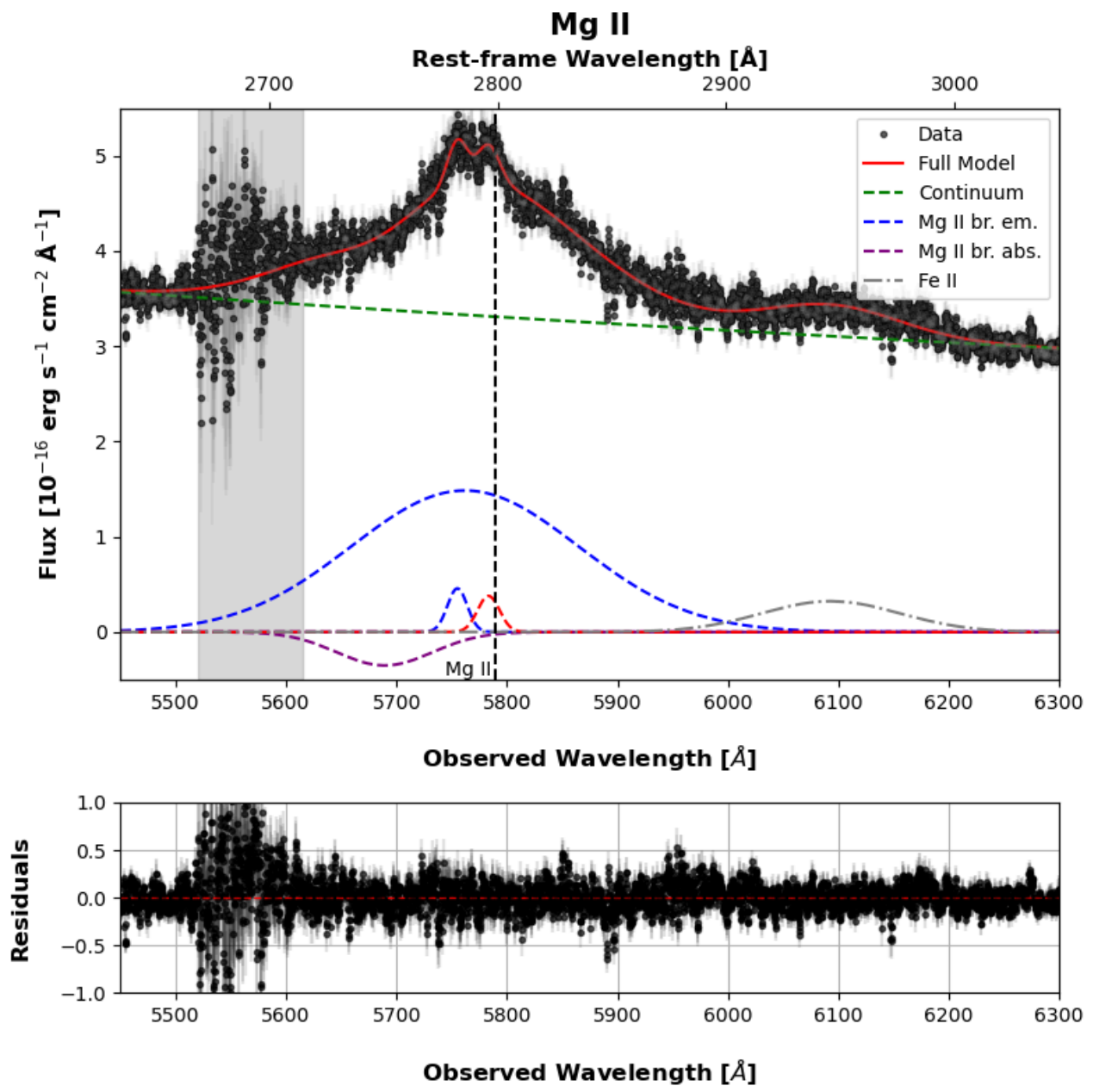

**Figure 6c. VLT/XSHOOTER spectrum of the Mg II emission line, with best fit model.** Same as for Fig 6a, but for Mg II. Shaded gray areas indicate regions that were masked out during the fitting process because of enhanced noise. The vertical line indicate the rest frame central wavelength of the line.

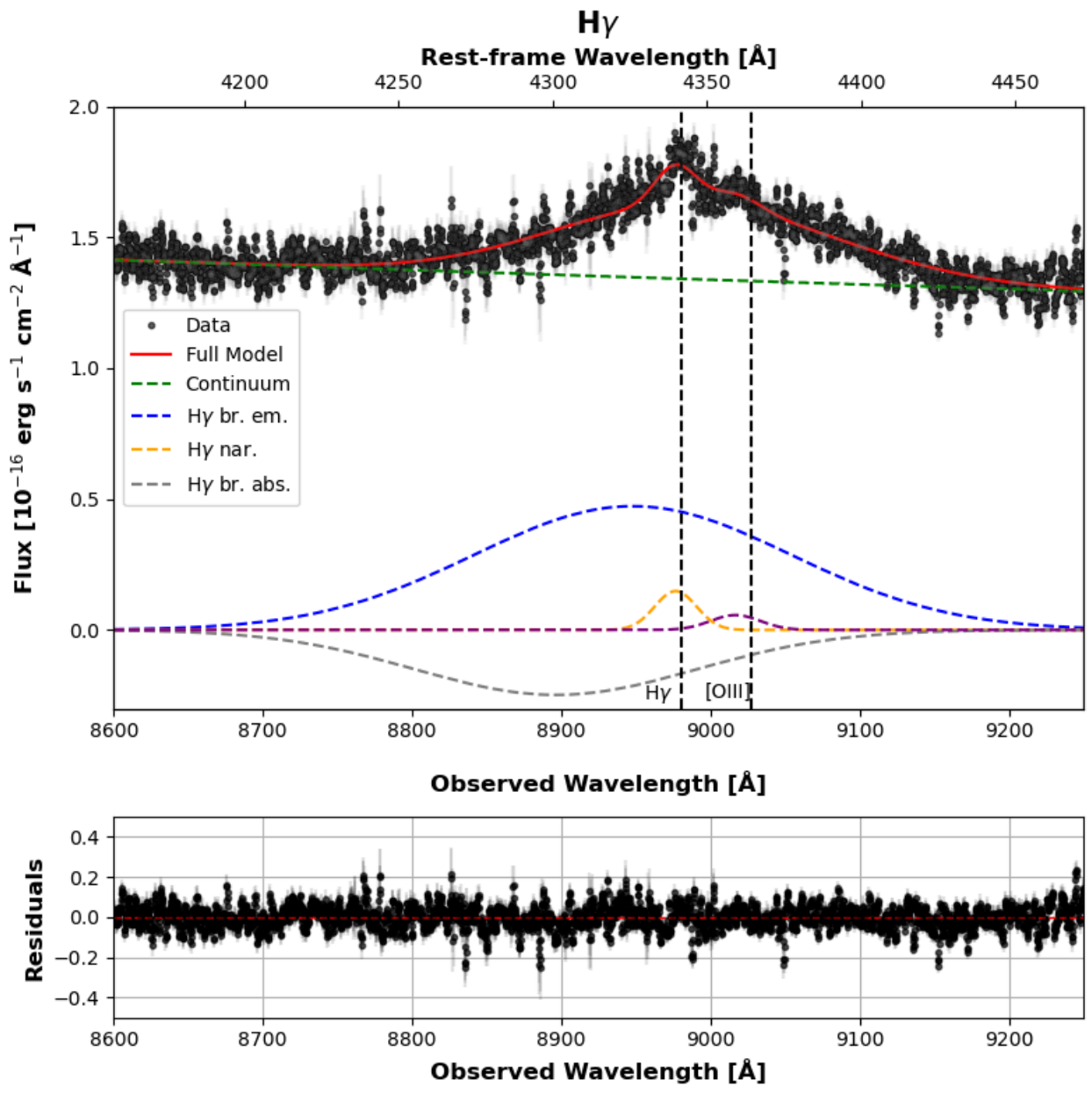

**Figure 6d VLT/XSHOOTER spectrum of the region of the Hγ emission line, with best fit model.** Same as for Fig 6a, but for Hγ and [OIII]4363. The vertical lines indicate the rest frame central wavelength of the lines.

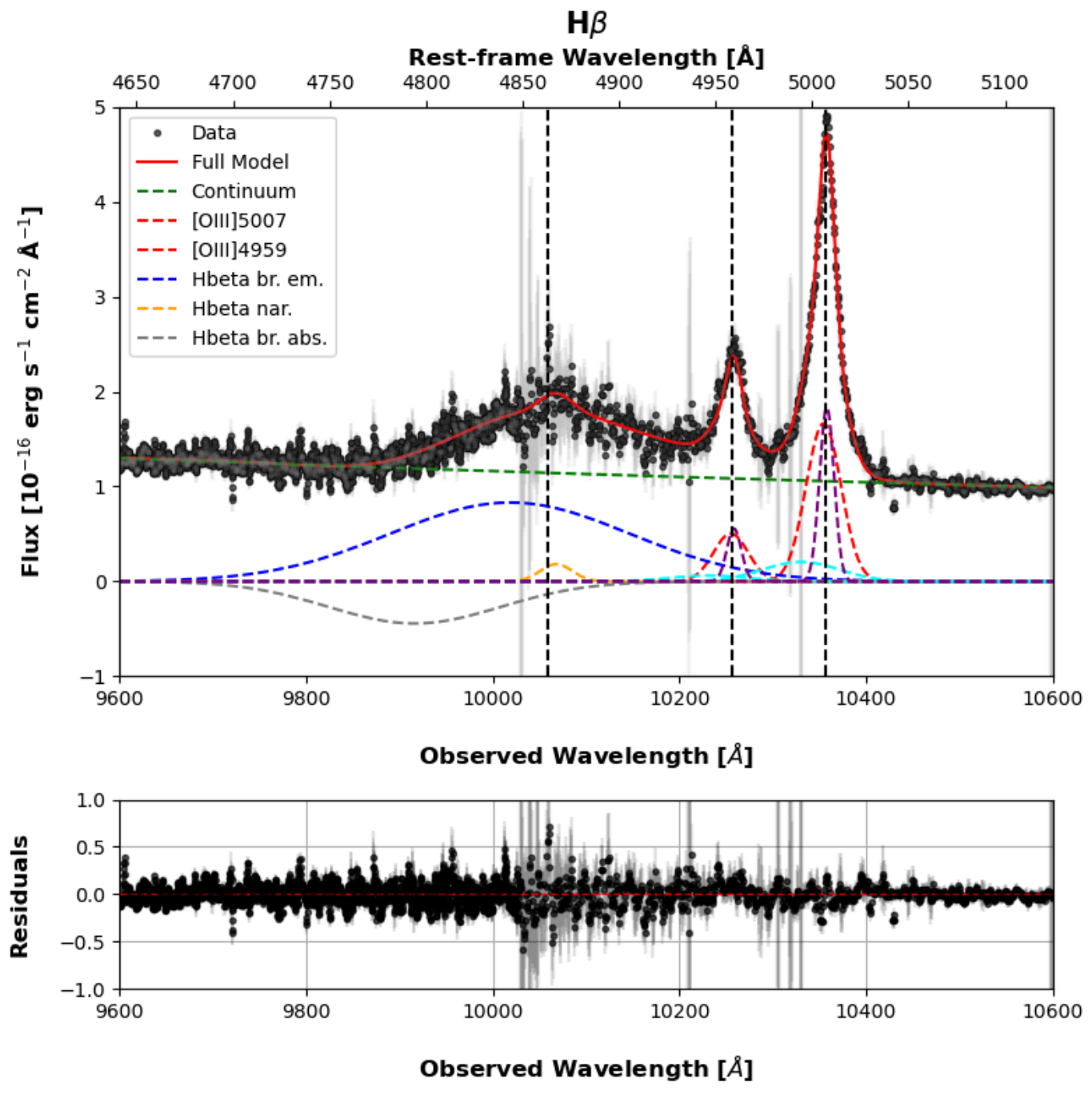

**Figure 6d VLT/XSHOOTER spectrum of the region of the Hβ + [OIII] complex, with best fit model.** Same as for Fig 6a, but for Hβ and [OIII]λλ5007,4959. The vertical lines indicate the rest frame central wavelength of the line.

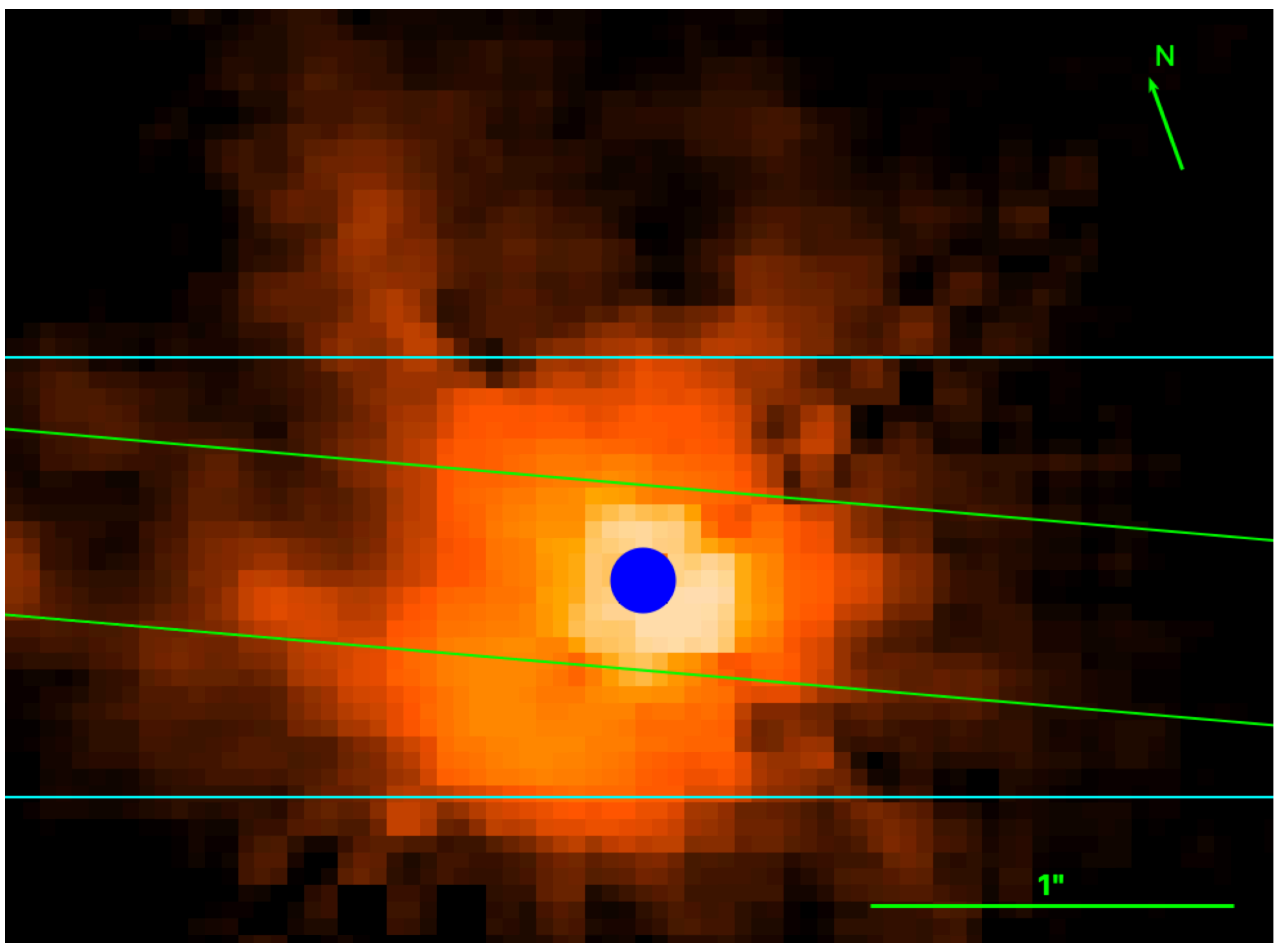

**Figure 7. Slit positions on the target.** The image is the same as Fig. 1c but with the slit position of the two datasets overlaid. The center of both slits is on the (PSF subtracted) QSO. The Subaru/SWIMS slit is marked in green and the VLT/XSHOOTER slit is in cyan. Note that the wider size of the VLT slit includes significantly more flux from the NLR, which explains the larger flux measured in some of the components of the [OIII] lines.

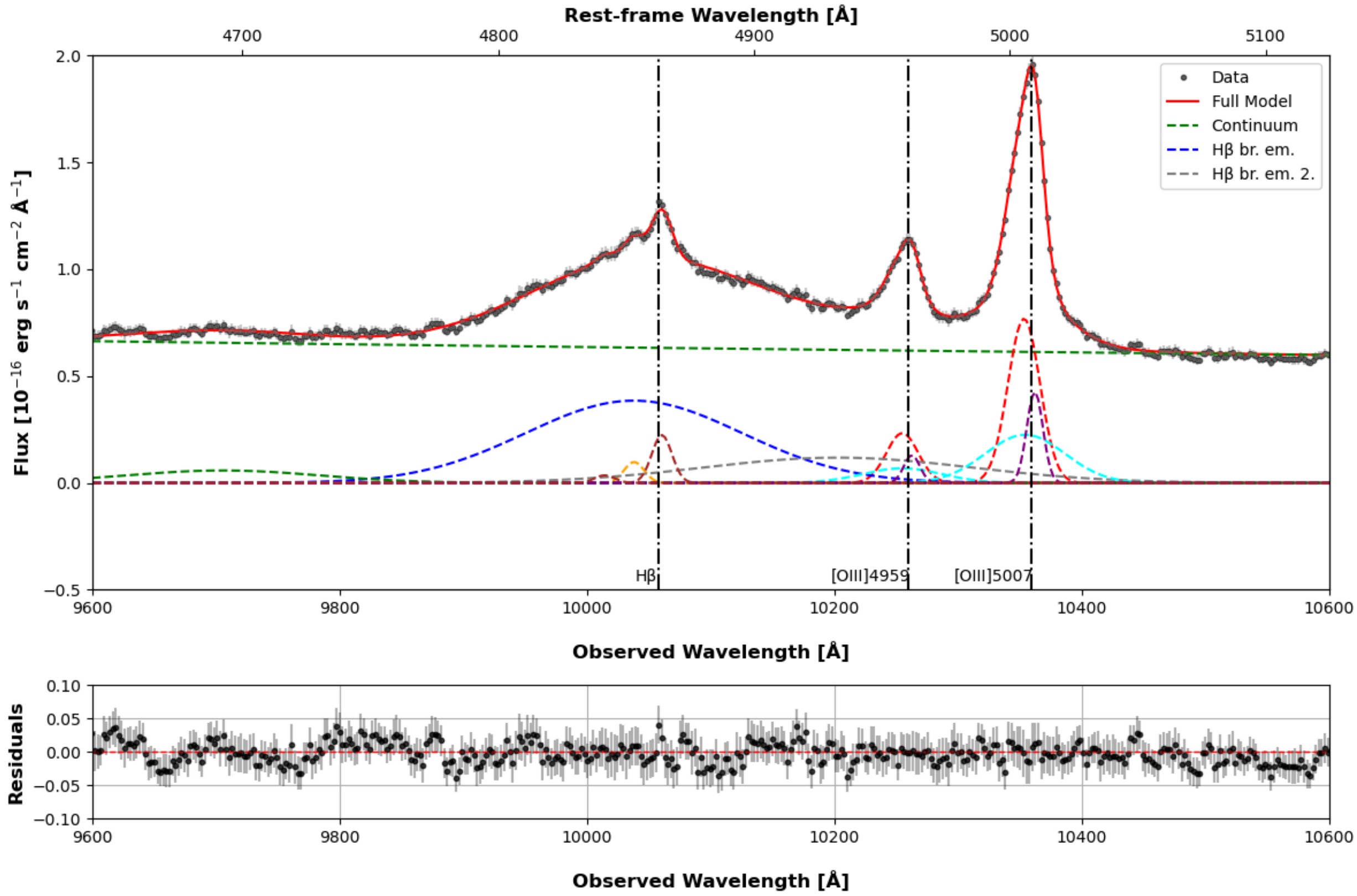

**Figure 8. Alternative model for Hβ (Subaru/SWIMS spectrum).** Same as for Figure 2 but using an alternative spectral model. The bottom x axis of the upper panel shows observed wavelengths and the top x axis shows the rest-frame wavelengths calculated assuming a redshift of $z_s = 1.0684$. Data points are black dots. Errorbars are at 2σ. The best fit model is shown with a red line on top of the data. Model components are shown as dashed lines: Hβ broad emission (two components, blue and gray, respectively), [OIII]λλ5007,4959 blueshifted, redshifted and broad-ish (FWHM ~ 2400km/s) components (purple, red and cyan, respectively, components a, b, and c in Table 1). The narrow components of Hβ are shown in orange and brown. The green dashed line is the power-law continuum. Residuals are shown in the bottom panel. Vertical dashed lines indicate the wavelengths of the relevant emission lines calculated for the systemic redshift $z_s = 1.0684$.

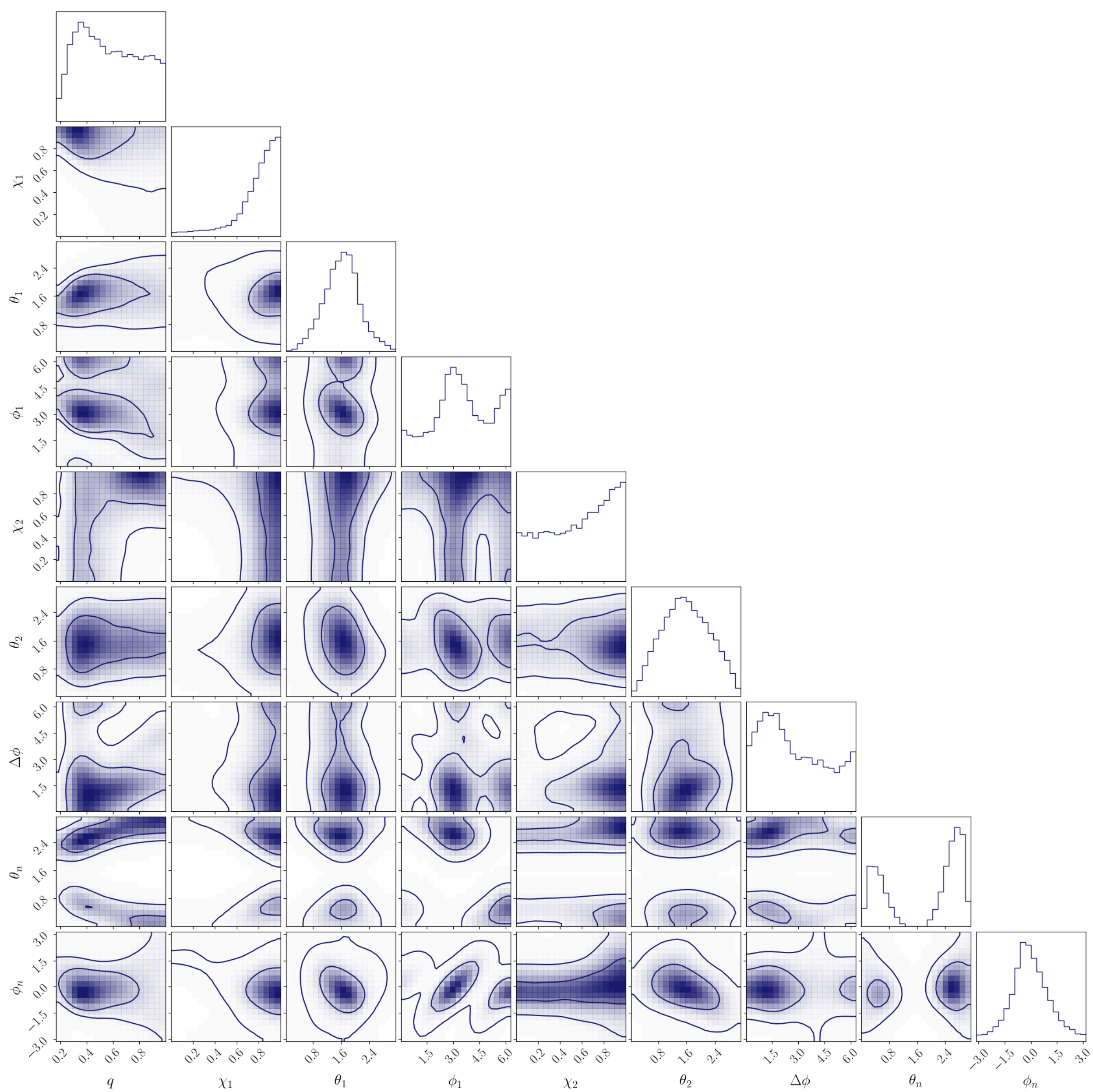

**Figure 9. Posterior distributions of the progenitor binary parameters**
One- and two- dimensional marginal distributions of all the sampled parameters: the mass ratio (q), the spin magnitudes ($\chi_{1,2}$), the spin tilt angles ($\theta_{1,2}$), the spin azimuthal angle of the heavier BH ($\phi_1$), the in-plane difference of the azimuthal angles ($\Delta\phi$), the observer polar angle ($\theta_n$), and the observer azimuthal angle ($\phi_n$). Contours refer to the 90% and 50% Bayesian credible regions.

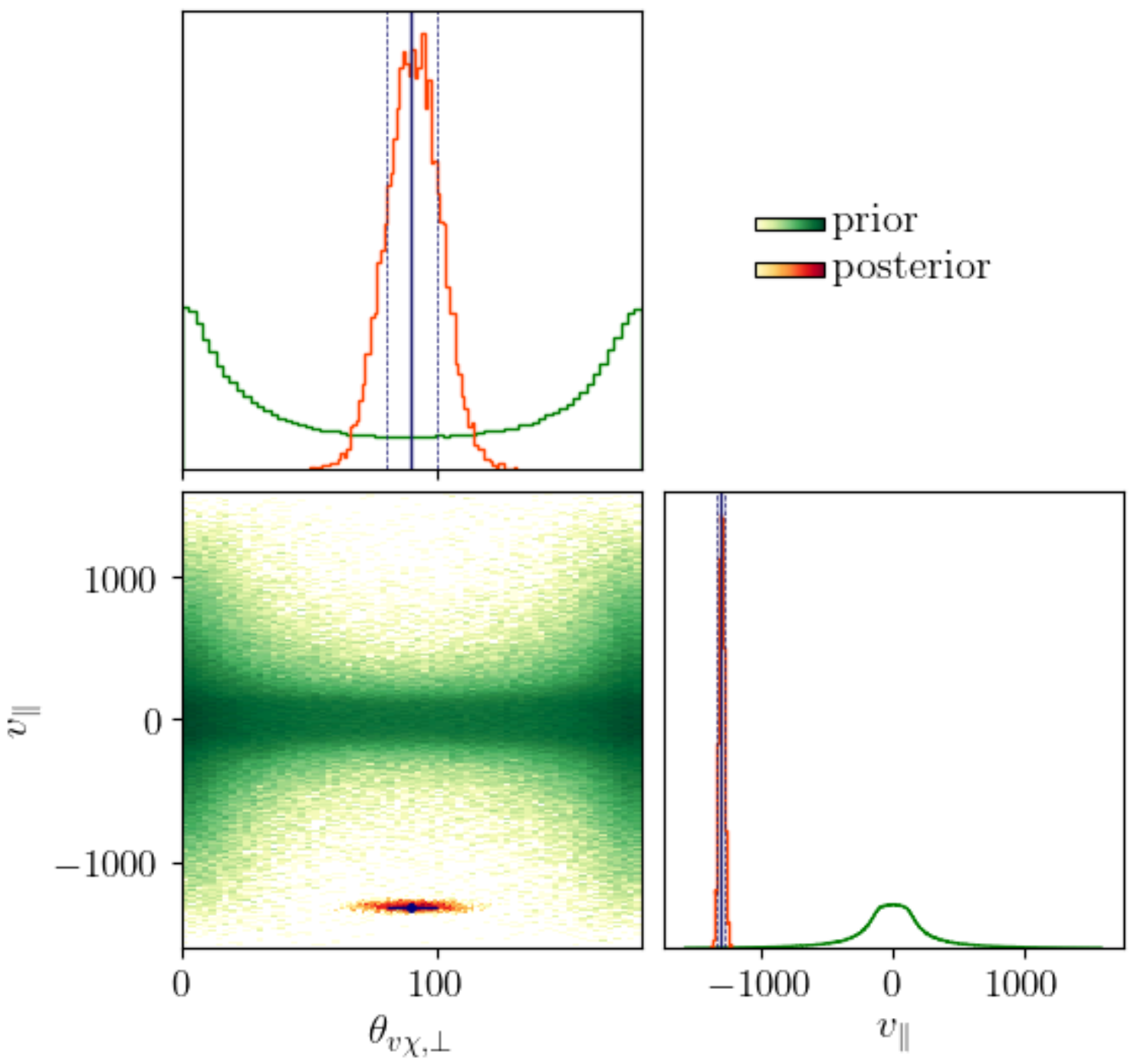

**Figure 10. One- and two- dimensional marginal prior and posterior distributions for parameters that directly enter the likelihood.**
Uninformative prior (green) and posterior distributions (red) resulting from our MCMC analysis. We consider the two quantities that directly enter the likelihood: the line-of-sight kick velocity ($v_\parallel$) and the angle between the spin and the kick projections perpendicular to the observer ($\theta_{v\chi,\perp}$). Higher (lower) probability density regions are shown in darker (lighter) shades. Blue lines and scatter points with error bars indicate means and standard deviations of our bivariate Gaussian likelihood distribution.

| Line model component | Observed Wavelength (Å) | Velocity Offset (km/s) | FWHM (km/s) | Line Flux ($10^{-16}$ erg s$^{-1}$ cm$^{-2}$) |
|---|---|---|---|---|
| Hβ Broad Emission | 10003.1 ± 8.4 | -1635 ± 253 | 10501 ± 333 | 174.0 ± 10.8 |
| Hβ Broad Absorption | 9894.2 ± 4.4 | -4881 ± 131 | 5916 ± 381 | -51.8 ± 9.9 |
| Hβ Narrow Em., a | 10060.8 ± 0.4 | +84 ± 11 | 570 ± 30 | 5.2 ± 0.3 |
| Hβ Narrow Em., b? | 10038.3 ± 1.0 | -571 ± 29 | 570 ± 30 | 2.6 ± 0.2 |
| [OIII]λ5007 Component a | 10353.2 ± 0.5 | -167 ± 14 | 882 ± 37 | 25.0 ± 2.3 |
| [OIII]λ5007 Component b | 10362.3 ± 0.4 | +95 ± 11 | 449 ± 27 | 6.9 ± 0.9 |
| [OIII]λ5007 Component c | 10353.8 ± 1.2 | -150 ± 34 | 2475 ± 250 | 20.0 ± 1.6 |

**Table 1: Best fit parameters for the Line Measurements (Subaru/SWIMS observations).** The model components for each line are listed in the first column (see Hβ+ [OIII] complex, for details). The table also reports the measured observed central wavelength for each component, the corresponding velocity offset with respect to the systematic redshift of the host galaxy $z_s$=1.0684, the FWHM in km/s of each component, and the integrated line flux.

| Line model component | Observed Wavelength [Å] | Velocity Offset [km/s] | FWHM [km/s] | Line Flux [$10^{-16}$ erg $s^{-1}$ $cm^{-2}$] |
|---|---|---|---|---|
| Hβ Broad Emission | 10018.6 ± 6.4 | -1173 ± 190 | 9021 ± 175 | 269.0 ± 18.8 |
| Hβ Broad Absorption | 9916.1 ± 5.3 | -4228 ± 158 | 6554 ± 142 | -102.7 ± 15.6 |
| Hβ Narrow Emission | 10067.8 ± 2.6 | +292 ± 66 | 1202 ± 183 | 7.6 ± 1.1 |
| [OIII]λ5007 Component a | 10354.1 ± 0.5 | -58 ± 14 | 1261 ± 61 | 74.4 ± 4.9 |
| [OIII]λ5007 Component b | 10357.8 ± 0.3 | +49 ± 9 | 537 ± 36 | 36.4 ± 5.3 |
| [OIII]λ5007 Component c | 10338.7 ± 2.4 | -503 ± 69 | 2805 ± 90 | 24.8 ± 4.9 |
| Hγ Broad Emission | 8947.9 ± 4.0 | -1082 ± 163 | 8390 ± 190 | 126.0 ± 2.9 |
| Hγ Broad Absorption | 8896.4 ± 3.7 | -2802 ± 123 | 7491 ± 41 | -58.7 ± 2.7 |
| Hγ Narrow Emission | 8976.7 ± 0.9 | -121 ± 30 | 1095 ± 277 | 5.1 ± 1.2 |
| [OIII]λ4363 | 9016.7 ± 3.7 | -355 ± 123 | 1303 ± 251 | 2.3 ± 0.2 |
| Mg II Broad Emission | 5761.9 ± 1.5 | -1352 ± 77 | 12580 ± 118 | 382.2 ± 6.0 |
| Mg II Broad Absorption | 5692.9 ± 2.2 | -4972 ± 114 | 5359 ± 301 | -38.0 ± 4.1 |
| Mg II Narrow, 1 | 5755.0 ± 0.8 | -1756 ± 41 | 1100 ± 103 | 10.2 ± 1.2 |
| Mg II Narrow, 2 | 5783.2 ± 1.1 | -295 ± 57 | 1212 ± 133 | 9.4 ± 1.0 |
| Fe II? | 6093.0 ± 1.9 | -1863 ± 39 | 7229 ± 106 | 50.7 ± 1.3 |
| CIII] Broad Emission | 3930.5 ± 0.37 | -1329 ± 28 | 8444 ± 93 | 192.2 ± 1.6 |
| CIII] Narrow Emission | 3942.6 ± 0.2 | -410 ± 22 | 1429 ± 68 | 12.3 ± 0.7 |
| Si II? | 3751.0 ± 9.0 | -574 ± 350 | 9381 ± 555 | 15.9 ± 2.3 |
| Al III? | 3821.0 ± 4.7 | -1623 ± 343 | 8566 ± 369 | 39.5 ± 1.3 |

**Table 2: Best fit parameters for the Line Measurements (VLT/XSHOOTER observations).** The model components for each line are listed in the first column (see text in Material and Methods, for details). The table also reports the measured observed central wavelength for each component, the corresponding velocity offset with respect to the systematic redshift of the host galaxy $z_s$=1.0684 the FWHM in km/s of each component, and the integrated line flux.

| Model parameter | Value |
|---|---|
| Mg II broad emission velocity | $-1363_{-117}^{+94}$ km/s |
| Mg II broad emission line flux | $3.58_{-0.05}^{+0.08}$ x $10^{-14}$ erg $s^{-1}$ $cm^{-2}$ |
| Mg II broad line FWHM | 12347 ± 126 km/s |
| Mg II broad absorption velocity | $-5321_{-90}^{+68}$ km/s |
| Fe II flux | (5.6 ± 0.1) x $10^{-15}$ erg $s^{-1}$ $cm^{-2}$ |
| Disk Velocity | $-1288_{-25}^{+29}$ km $s^{-1}$ |
| Disk Inclination | $<$ 9 deg |
| $r_{in}$ | 820 ± 20 $r_g$ |
| $r_{out}$ | 2000 ± 300 $r_g$ |

**Table 3: Mg II line *KERRDISK* model parameters.** Relevant model parameters for the best fit to the Mg II emission line complex using the *KERRDISK* accretion disk model. Values are given with units. Inner and outer radii ($R_{in}$ and $R_{out}$) are given in units of the gravitational radius $r_G = GM/c^2$.

| Line model component | Observed Wavelength [Å] | Velocity Offset [km/s] | FWHM [km/s] | Line Flux [$10^{-16}$ erg $s^{-1}$ $cm^{-2}$] |
|---|---|---|---|---|
| Hβ Broad Emission | 10037.8 ± 1.9 | -598 ± 57 | 6146 ± 139 | 84.3 ± 2.2 |
| Broad comp. 2 | 10202.7 ± 12.0 | +4316 ± 359 | 7544 ± 438 | 32.3 ± 4.3 |
| Hβ Narrow Em., a | 10060.5 ± 0.4 | +75 ± 11 | 569 ± 30 | 4.5 ± 0.3 |
| Hβ Narrow Em., b? | 10038.0 ± 1.1 | -601 ± 32 | 569 ± 30 | 2.0 ± 0.2 |

**Table 4: alternative model for Hβ.** The model components for each line are listed in the first column (see Alternative model for Hβ, for details). The table also reports the measured observed central wavelength for each component, the corresponding velocity offset with respect to the systematic redshift of the host galaxy $z_s$=1.0684 the FWHM in km/s of each component, and the integrated line flux.

| Line | Observed Flux $F_\lambda$ [erg cm^-2 s^-1 A^-1] | Luminosity $\lambda L_\lambda$ [erg s^-1] | FWHM [km/s] | $M_{BH}$ ($M_{sun}$) |
|---|---|---|---|---|
| Hβ | (5.84 ± 0.01) x $10^{-17}$ at 10548 Å | 1.8 x $10^{45}$ at 5100 Å | 10501 ± 333 | 2.4 x $10^9$ |
| Mg II | (3.0 ± 0.1) x $10^{-16}$ at 6205 Å | 5.6 x $10^{45}$ at 3000 Å | 12580 ± 118 | 4.7 x $10^9$ |

**Table 5: Black hole mass estimation.** The broad emission line used for each BH estimate is listed in column (1), the continuum flux is in column (2), the derived luminosity, at the distance corresponding to $z_s$=1.0684 is in column (3), the FWHM of the line is in column (4) and the derived BH mass using appropriate relations [60] is in column (5).

| | $0° < \theta_n < 90°$ | $90° < \theta_n < 180°$ |
|---|---|---|
| q | $0.58_{-0.32}^{+0.37}$ | $0.54_{-0.30}^{+0.41}$ |
| $\chi_1$ | $0.83_{-0.57}^{+0.16}$ | $0.84_{-0.52}^{+0.14}$ |
| $\theta_1$ [deg] | $95.9_{-50.7}^{+47.7}$ | $90.8_{-48.3}^{+49.0}$ |
| $\phi_1$ [deg] | $-28.1_{-94.8}^{+91.3}$ | $171.4_{-102.6}^{+90.7}$ |
| $\chi_2$ | $0.64_{-0.58}^{+0.33}$ | $0.61_{-0.55}^{+0.36}$ |
| $\theta_2$ [deg] | $93.1_{-61.6}^{+61.1}$ | $87.0_{-60.0}^{+68.2}$ |
| $\Delta\phi$ [deg] | $135.1_{-116.1}^{+204.6}$ | $131.7_{-115.0}^{+209.4}$ |
| $\theta_n$ [deg] | $25.4_{-17.7}^{+28.7}$ | $153.1_{-28.5}^{+18.0}$ |
| $\phi_n$ [deg] | $-17.3_{-98.8}^{+130.0}$ | $1.0_{-105.1}^{+110.6}$ |
| $m_f$ [M] | $0.96_{-0.02}^{+0.01}$ | $0.96_{-0.02}^{+0.01}$ |
| $\chi_f$ | $0.68_{-0.13}^{+0.13}$ | $0.70_{-0.14}^{+0.14}$ |
| v [km/s] | $1328_{-42}^{+62}$ | $1328_{-42}^{+59}$ |
| $v_{\parallel}$ [km/s] | $-1309_{-35}^{+34}$ | $-1309_{-34}^{+35}$ |
| $v_{\perp}$ [km/s] | $195_{-145}^{+248}$ | $195_{-141}^{+235}$ |
| $\theta_{v\chi,\perp}$ [deg] | $87.0_{-16.8}^{+17.0}$ | $92.9_{-17.2}^{+16.7}$ |
| $\theta_{v\chi}$ [deg] | $12.9_{-7.4}^{+10.3}$ | $166.9_{-10.1}^{+7.3}$ |
| $\theta_{n\chi}$ [deg] | $8.6_{-6.4}^{+11.1}$ | $171.2_{-11.0}^{+6.5}$ |
| $\theta_{nv}$ [deg] | $8.5_{-6.3}^{+10.2}$ | $8.5_{-6.1}^{+9.8}$ |
| $\theta_{jet}$ [deg] | $8.6_{-6.4}^{+11.1}$ | $8.8_{-6.5}^{+11.0}$ |

**Table 6: medians and 90% credible intervals of the marginalized distributions**

We report medians and the 90% symmetric credible intervals for the marginal distributions of some parameters entering this analysis. We separate two different modes, corresponding to configurations where the recoil velocity is either aligned or anti-aligned with respect to the BH spin. The first set of parameters enter directly our sampling and refers to the progenitor binary properties and the observer line of sight: mass ratio (q), spin magnitudes ($\chi_{1,2}$), spin tilt angles ($\theta_{1,2}$), spin azimuthal angle of the heavier BH ($\phi_1$), in-plane difference of the azimuthal angles ($\Delta\phi$), observer polar angle ($\theta_n$), and observer azimuthal angle ($\phi_n$). We also build posteriors of remnant properties, such as the final mass ($m_f$) in total mass unit, the remnant spin ($\chi_f$) and the kick (v) magnitudes. Finally, we present projected quantities and angles, some of which directly enter our likelihood: the projected kick velocity ($v_{\parallel}$), the angle between the sky-projected kick and spin ($\theta_{v\chi,\perp}$), the velocity component projected onto the sky ($v_\perp$), the kick-spin angle ($\theta_{v\chi}$), the observer-spin angle ($\theta_{n\chi}$), the observer-kick angle ($\theta_{nv}$), and the jet viewing angle ($\theta_{jet}$).

**Data availability**

The data in the manuscript and Supplementary materials are available for download through public archives. For HST, data are in the Mikulsky Archive for Space Telescopes (MAST) at http://mast.stsci.edu. The VLT/XSHOOTER spectrum can be retrieved from the ESO archive at http://archive.eso.org. Subaru/SWIMS data are available from the Subaru data archive at http://smoka.nao.ac.jp/.

**Acknowledgments**

MC wishes to thank Arieh Königl for suggestions on the timescales of the GW event and the AGN activity, and Konstantinos Kritos, Emanuele Berti, and Julian Krolik for helpful comments and suggestions. The authors thank Bryan Hilbert for providing a copy of the PSF-subtracted HST images. M.B. and D.G. are supported by ERC Starting Grant No. 945155-GWmining, Cariplo Foundation Grant No. 2021-0555, MUR PRIN Grant No. 2022-Z9X4XS, MUR Grant "Progetto Dipartimenti di Eccellenza 2023-2027" (BiCoQ), and the ICSC National Research Centre funded by NextGenerationEU. D.G. is supported by MSCA Fellowships No. 101064542-StochRewind and No. 101149270-ProtoBH. Computational work was performed at CINECA with allocations through INFN and Bicocca. G.C. acknowledges the support from the Next Generation EU funds within the National Recovery and Resilience Plan (PNRR), Mission 4 - Education and Research, Component 2 - From Research to Business (M4C2), Investment Line 3.1 - Strengthening and creation of Research Infrastructures, Project IR0000012 – "CTA+ - Cherenkov Telescope Array Plus".

Based on observations collected at the European Southern Observatory under ESO programme 110.23WJ https://doi.eso.org/10.18727/archive/71
This research is based on observations made with the NASA/ESA *Hubble Space Telescope* obtained from the Space Telescope Science Institute, which is operated by the Association of Universities for Research in Astronomy, Inc., under NASA contract NAS 5–26555. These observations are associated with program 15254.

**Author contributions**

MC wrote the VLT proposal, prepared the observations, reduced spectra (VLT), analyzed all spectra performing spectral fits with *specfit,* derived results, made figures, provided the basic interpretation of the results, calculated the timescales and wrote most of the manuscript. TM wrote the Subaru/SWIMS proposal, contributed to the SWIMS data reduction and processing of the HST images, and wrote the corresponding section in the manuscript. MB ran the numerical relativity model, wrote the corresponding sections in the manuscript. SB performed the spectral fit using the *KERRDISK* model. AC discussed the interpretation of the results and participated in the writing of the manuscript and the VLT observing proposal. CN contributed to the interpretation of the results and the calculation of the relevant timescales. MS contributed to the interpretation of the results, and to the discussion of different spectral models and resulting emission line ratios. GC, EL, EM, and GT provided comments to the manuscript and contributed to the interpretation of the results. KM, SK, HT, MK planned and performed the SWIMS observations, and contributed to the data reduction. KK contributed to the SWIMS data reduction. All co-authors discussed the manuscript and provided comments.